\documentclass[12pt,a4,onecolumn,draftclsnofoot,oneside, journal,compsoc]{IEEEtran}

\usepackage{amssymb}
\usepackage{amsmath}
\usepackage{amsfonts}
\usepackage{graphicx}
\usepackage{subfigure}
\usepackage[nocompress]{cite}
\usepackage{array}
\usepackage{enumerate}
\usepackage{url}

\begin{document}

\title{{Collaborative Smartphone Sensing using Overlapping Coalition Formation Games}}
\author{
\IEEEauthorblockN{
\normalsize{Boya Di}\IEEEauthorrefmark{1},
\normalsize{Tianyu Wang}\IEEEauthorrefmark{1},
\normalsize{Lingyang Song}\IEEEauthorrefmark{1},
\normalsize{and Zhu Han}\IEEEauthorrefmark{2},\\}
\IEEEauthorblockA{\small
\IEEEauthorrefmark{1}School of Electrical Engineering and Computer Science, Peking University, Beijing, China. \\
\IEEEauthorrefmark{2}Electrical and Computer Engineering Department, University of Houston, Houston, TX, USA.} \\
}

\maketitle
\begin{abstract}

With the rapid growth of sensor technology, smartphone sensing has become an effective approach to improve the quality of smartphone applications. However, due to time-varying wireless channels and lack of incentives for the users to participate, the quality and quantity of the data uploaded by the smartphone users are not always satisfying. In this paper, we consider a smartphone sensing system in which a platform publicizes multiple tasks, and the smartphone users choose a set of tasks to participate in. In the traditional non-cooperative approach with incentives, each smartphone user gets rewards from the platform as an independent individual and the limit of the wireless channel resources is often omitted. To tackle this problem, we introduce a novel cooperative approach with an overlapping coalition formation game (OCF-game) model, in which the smartphone users can cooperate with each other to form the overlapping coalitions for different sensing tasks. We also utilize a centralized case to describe the upper bound of the system sensing performance. Simulation results show that the cooperative approach achieves a better performance than the non-cooperative one in various situations.

\end{abstract}


\newpage
\section{Introduction}%
Smartphones today, with their good programmability and various embedded sensors, are no longer just communication devices. They can also monitor and collect data from surroundings, which then drive the development of new research on smartphone sensing. The smartphone users can upload their sensed data for further processing and help identify the routines of target population. Plenty of new applications have been proposed based on smartphone sensing, e.g., traffic monitoring~\cite{8MM-2010}\cite{19TRLMBTE-2009}, healthcare~\cite{16OM-2007}\cite{24DACHLBD-2009}, and even social networking~\cite{21BC-2010}. Generally, these applications can be classified into three types~\cite{3LMLPCC-2010}: individual sensing, group sensing and community sensing. We mainly focus on the community sensing~\cite{5RFH-2011}~\cite{15J-2012}, also known as mobile crowdsourcing in the narrow sense, in which a certain number of users are encouraged to participate in the smartphone sensing to ensure the quality of the large-scale applications.

Typically, a smartphone sensing system comprises a sensing platform with several back-end servers on the Internet, and many smartphone users embedded with various sensors. The platform publicizes multiple tasks, and recruits the smartphone users to provide sensing services. The users then select and participate in one or more tasks, and upload the sensed data to the platform for further analysis. When sensing and uploading the data, the users need to consume their own resources, such as power, memory, time, and wireless channel resources. Therefore, the users may not be interested in participating in the sensing tasks, unless they are rewarded to compensate their consumption of resources. The sensing performance of the tasks cannot be guaranteed if the platform do not recruit enough users, which then affects the quality of services that the platform can provide. Thus, the incentive mechanisms for motivating the smartphone users need to be considered~\cite{JVL-2012}\cite{2DGXJ-2012}.

Some work has noticed the problem mentioned above and provides different incentive mechanisms for the smartphone sensing systems. In~\cite{2DGXJ-2012}, the users introduced a platform-centric model in which the platform publicizes only one task, and a user-centric model in which each user asks for a price for its sensed data. The user-centric model is studied as a matching market and the truthfulness of pricing is guaranteed. In~\cite{25XCW-2009}, the authors proposed a bargain-based mechanism to encourage the cooperative message trading among the selfish nodes to maximize their rewards. Each task can only be completed by one user, and they model the message transaction as a two-person cooperative game, in which the Parieto optimum is achieved. In~\cite{6LH-2010} -\cite{4THL-2014}, different incentive mechanisms based on various branches of the auction theory are designed to reach a balance between the quality of the sensing services and the incentive cost.

Remarkably, most researchers assume that the users make independent decisions when choosing the sensing tasks, and they do not know how their strategies influence the sensing performance of each task. Based on this assumption, it is quite likely that most users tend to participate in a popular task, while the other tasks cannot recruit enough users. This may result in an unequal distribution of the users' resources, and thus, both the quality of the sensing services and the users' rewards are affected. Besides, in most works, the authors have not considered the limitation of each user's available wireless channel resources. However, in a general case in which the tasks require real-time data uploading, or the required data feedback rate for uploading is relatively high, the users' limited resources may influence the sensing performance of the smartphone sensing networks.

To avoid the resource imbalance problem mentioned above, we consider the cooperation among the users. Through cooperating and exchanging information with each other, the users can know the specific situation of their preferred tasks so as to make wiser choices of resource allocation. Taking into account the selfishness and rationality of the users, we regard coalitional game theory as a suitable mathematical tool for modeling the user cooperation and the internal relationship between them\cite{9HNSBH-2011} -\cite{27SWKC-2012}. However, we find that most works utilizing the coalition formation games for the user cooperation assume that one user can only join one coalition\cite{32BM-2013}\cite{28BM-2012} -\cite{FD-2007}, which does not quite fit our scenario. In the smartphone sensing network, since each user can be involved in multiple tasks, so the coalitions representing different sensing tasks may overlapping with each other.

To capture this characteristic of the network, we then introduce the overlapping coalition formation game (OCF-game)~\cite{1CEMJ-2010} in which the rational players can simultaneously join multiple coalitions. Several works have applied the OCF-game in various fields~\cite{11ZSHSL-2013} -\cite{42XPD-2014}. In~\cite{11ZSHSL-2013}, the authors considered both each user's personal profits and how they influence the social welfare in a small-cell network. In~\cite{22DDRJ-2006}, the authors assume that the users are fully cooperative and the maximum social welfare can be obtained. In~\cite{42XPD-2014}, OCF-games are utilized to model the cooperation between the service providers in the wireless relay networks, and a brief remark on the stability of a merge-and-split algorithm is given. Though the assumptions and algorithms in these works are suitable in the specific scenarios, few works have paid enough attention to the fact that most users are rational players who only aim at maximizing their own welfare. In addition, the information exchange and the cooperation cost should be considered when we model an OCF-game.

The main contribution of this paper can be summed up as below. We aim at designing an incentive mechanism in which the platform encourages the smartphone users to participate in the sensing tasks by offering them rewards. In this mechanism, we consider a general case in which the limited wireless channel resources may affect the quality of the sensing services. To improve the sensing performance of the platform, we adopt a novel cooperative approach in which the users cooperate to form different coalitions based on various tasks, and the platform gives rewards directly to the coalitions, not the users. We formulate the task selection problem as an OCF-game and we propose a distributed overlapping coalition formation algorithm (OCF-algorithm) in which the users can maximize their own profits by selecting multiple coalitions to join and investing the wireless resources. The properties of our proposed OCF-algorithm are then analyzed. We also propose a traditional non-cooperative incentive mechanism which serves as a comparison with the cooperative approach. Simulation results show that the proposed cooperative approach with the OCF-algorithm can perform better than the non-cooperative approach.

The rest of this paper is organized as follows. In Section \textbf{2}, we provide the system model of both the non-cooperative approach and the cooperative approach. In Section \textbf{3}, we formulate the non-cooperative approach as a Stackelberg game, and solve it utilizing the nonlinear optimization theory. In Section \textbf{4}, we formulate the cooperative approach as an OCF-game and propose an OCF-algorithm. In Section \textbf{5}, simulation results are presented and analyzed. Finally, we conclude the paper in Section \textbf{6}.

\section{System Model}%
Consider a smartphone sensing system consisting of a platform with several servers on the Internet, one wireless base station (BS), and some smartphones embedded with various sensors, as shown in Fig.~1. In the system, the sensing platform publicizes $N$ sensing tasks, the set of which is denoted by ${\cal N} = \{ 1, \cdots ,N\}$. The set of $M$ smartphone users, denoted by $\mathcal{M} = \{ 1, \cdots ,M\}$, can select multiple tasks to participate in and then upload their sensed data to the platform via the BS. Without loss of generality, we assume that the BS allocates $K$ non-overlapping subcarriers, the set of which is denoted by $\mathcal{K} = \{1, \cdots, K\}$, to the smartphone users in a single-cell OFDMA network. The users then need to invest a certain amount of wireless channel resources into their participating tasks for uploading their sensed data. We assume that each task has a central location\footnote{This assumption is valid in practice. For example, if there is a task aiming at getting the noise map of a certain area of interest, then we can set the centre of this area as the central location of this task.}. All the tasks and the smartphone users are randomly spread over a square area with the side length $L$.
\begin{figure}[!t]
\centering
\includegraphics[width=3in]{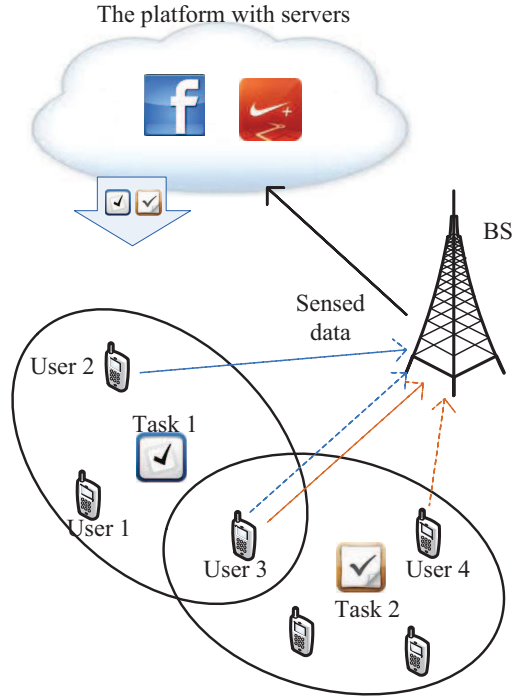}
\caption{System model of the smartphone sensing system.} \label{system_model}
\end{figure}

We define a $K \times M$ capacity matrix $\emph{\textbf{C}} = {\left[ {{c_{i,j}}} \right]_{K \times M}}$, where ${c_{k,j}}$ is the capacity of subcarrier $k$ for user $j$. The element $c_{k,j}$ can be given by
\begin{align} \label{capacity_define}
{c_{k,j}} = B{\log _2}\left( {1 + \frac{{{p_s}{{\left| {{h_{k,j}}} \right|}^2}}}{{{\sigma _n}^2}}} \right),
\end{align}
where $B$ is the bandwidth of each subcarrier, $p_s$ is the transmitted power of the BS to each user, ${h_{k,j}} \sim \mathcal{CN}\left( {0,{D_j}^{ - \delta }} \right)$ is the Rayleigh channel, $\delta $ is the path loss exponent, and ${{\sigma _n}^2}$ is the noise variance.

To better describe the task selection of the users and the subcarrier allocation of the BS, we define a $N \times M$ task matrix $\emph{\textbf{X}} = {\{ 0,1\}}$, in which ${x_{i,j}} = 1$ denotes that user $j$ participates in task $i$, and a $K \times M$ subcarrier matrix $\textsl{\textbf{S}} = {\{ 0,1\}}$, where ${s_{k,j}} = 1$ denotes that subcarrier $k$ is assigned to user $j$.

It is worth noting that some tasks may require the participating users to upload quite a bit of real-time sensed data, or the channel condition between the BS and a user may not be satisfying. Therefore, it is possible that a user cannot finish the tasks due to its limited data feedback rate. For any task $i \in \cal{N}$, we assume that the minimum resources that a participating user $j$ needs to invest in task $i$ is set to be data feedback rate ${r_i}$. Thus, we have the bandwidth constraint of each user as:
\begin{align} \label{system_constraint1}
{\sum\limits_{k \in \cal{K}}{{s_{k,j}}{c_{k,j}}}  \ge \sum\limits_{i \in \cal{N}} {{r_i}{x_{i,j}}} },\forall j \in {\cal M}.
\end{align}
For those cases in which the tasks require low feedback rate and the channel condition is good enough, the bandwidth constraint is always satisfied.  The system model can then be degraded to one without bandwidth constraint as shown in \cite{32my-2013}. Since each subcarrier can only be assigned to one user, $\textbf{\emph{S}}$ is required to satisfy:
\begin{align} \label{system_constraint2}
\sum\limits_{j \in \cal{M}} {{s_{k,j}}}  \le 1,\forall k \in {\cal K}.
\end{align}

For any user $j$ participating in task $i$, we assume that user $j$ is charged by the BS for the use of data feedback rate ${r_i}$ \footnote{Note that this charge is usually asked by the service providers who lease the BS.}. Thus, for any task $i \in {\cal{N}}$, the revenue that the BS gets from user $j$ is:
\begin{align} \label{charge_of_rate}
{R_{i,j}}(\emph{\textbf{X}}) = \beta {r_i}{x_{i,j}},
\end{align}
where $\beta$ is a scaling factor.

Once user $j$ participates in task $i$, it collects and uploads the sensed data that task $i$ requires. We assume that each task has a specific area of interest (AoI), i.e., the area from which a task collects the sensed data. User $j$'s contribution to task $i$ is related to the distance between them. The users in this area contribute the same to this task, while those outsiders contribute less than the insiders. Specifically, for any user $j \in {\cal{M}}$ and any task $i \in {\cal{N}}$, if user $j$ participates in task $i$, we define the contribution of user $j$ to task $i$ as:
\begin{align} \label{oneuser_onetask_utility}
{Q_{i,j}} = \left\{ \begin{array}{l}
\frac{{{a_i}}}{{{d_{i,0}}^\lambda }},{d_{i,j}} \le {d_{i,0}}\\
\frac{{{a_i}}}{{{d_{i,j}}^\lambda }},{d_{i,j}} > {d_{i,0}}
\end{array} \right.,
\end{align}
where $\lambda$ is an exponential factor, and ${{d_{i,0}}}$ is a constant representing the radius of task $i$'s AoI. Since different tasks are provided by various third-party companies, the contribution made to them is valued in different ways. One user may make different contribution to two different tasks even when the distances between the user and these two tasks are the same. Thus, we set $a_i$ as a scaling factor to describe the difference of the tasks.

We assume that the sensing performance of task $i$, also known as task $i$'s profit, increases proportionally with the total contribution of all the participating users, until the contribution reaches a threshold $\rho_i$ \footnote{This assumption makes sense. For example, there is a task which is to measure the noise map of a certain area, then once the number of the smartphone users who participate in this task exceed, say, 1000, those additional users make little contribution.}. Therefore, the sensing performance of any task $i$ is given by
\begin{equation} \label{onetask_performance}
{\Gamma_i}(\emph{\textbf{X}}) =
\begin{cases}
\frac{{\varphi_i}}{{\rho_i}} \sum\limits_{j \in \cal{M}} {{Q_{i,j}} {x_{i,j}}} ,  &~ \sum\limits_{j \in \cal{M}} {{Q_{i,j}} {x_{i,j}}}  \le \rho_i, \\
\varphi_i,  &~ \sum\limits_{j \in \cal{M}} {{Q_{i,j}} {x_{i,j}}}  > \rho_i,
\end{cases}
\end{equation}
where $\sum\nolimits_{j \in {\cal{M}}} {{Q_{i,j}} {x_{i,j}}}$ represents the total contribution that the users make to task $i$, and $\varphi_i$ is the upper bound of task $i$'s sensing performance which corresponds to $\rho_i$. The total profits of all the tasks, i.e., the platform sensing performance, is then given by:
\begin{align} \label{platform_sensing_performance}
{PFM} = \sum\limits_{i \in {\cal{N}}} {\Gamma_i(\emph{\textbf{X}})}.
\end{align}

Note that the users are not willing to participate in the tasks without getting paid. Therefore, the platform needs to reward the users for their contribution, which is usually called the incentive mechanism. For a practical smartphone sensing system, the BS allocates the subcarriers to the users and the platform provides an incentive mechanism, while the users try to maximize their own profits obtained from the platform. In the rest of this section, a centralized case is provided to be set as an upper bound, in which the users are fully scheduled by the platform and the BS. We then study a practical approach in which the users behave cooperatively, and set a non-cooperative approach as a benchmark.

\subsection{Centralized case}%
In the centralized case, the users are forced to participate in the tasks and get no rewards. The platform and the BS together decide how to assign the subcarriers to the users and how the users participate in the tasks. This case describes the upper bound of the system sensing performance which does not exist in practice, because it is impossible that the users are willing to participate in the tasks without any rewards.

The platform utility consists of two parts: the platform's sensing performance and part of the revenue from the BS. We present the platform utility as:
\begin{align} \label{platform_centric_utility}
{U_{CE}} = {PFM} + \sum\limits_{i \in {\cal{N}},j \in {\cal{M}}} \gamma {{R_{i,j}}} (\emph{\textbf{X}}),
\end{align}
where $\gamma$ is the scaling factor.

Given the constraints $\left( \ref{system_constraint1} \right)$ $\left( \ref{system_constraint2} \right)$, the centralized case is then formulated as:

\begin{align} \label{platform_centric_utility_1}
\begin{split}
& \mathop {\max }\limits_{\emph{\textbf{X}},\emph{\textbf{S}}} \left[ {\sum\limits_{i \in \cal{N}} {\Gamma_i} (\emph{\textbf{X}}) + \gamma \beta \sum\limits_{i \in \cal{N}} {{r_i}} \sum\limits_{j \in \cal{M}} {{x_{i,j}}} } \right] \\
& s.t.    \left\{ \begin{array}{l}
\sum\limits_{k \in {\cal{K}}} {{s_{k,j}}{c_{k,j}}}  \ge \sum\limits_{i \in {\cal{N}}} {{r_i}{x_{i,j}}}, \\
\sum\limits_{j \in {\cal{M}}} {{s_{k,j}} \le 1}.
\end{array} \right.
\end{split}
\end{align}
This is a 0-1 integer nonlinear programming problem (INLP problem), which can be approximated by a convex function and solved by utilizing an optimization algorithm \cite{12RR-2009}\cite{29BV-2004}.
\subsection{Non-cooperative approach}%
In the non-cooperative approach, the BS allocates the limited subcarriers to the users, and the platform rewards the users for their individual contribution to the tasks. The users make independent decisions on which tasks to participate in, based on the potential rewards they may get from the platform. For any user $j \in {\cal{M}}$, the rewards it gets are proportional to its total contribution to all the tasks, i.e., ${\alpha _1}\sum\nolimits_{i \in {\cal{N}}} {{Q_{i,j}} {x_{i,j}}} $, where ${\alpha_1}$ is a scaling factor describing the incentive intensity.

In practice, the service providers who lease the BS usually distribute part of the charge of data feedback rate to the platform as revenue splits (for example, T-Mobile and the third-party applications on T-Mobile Partner Network, PCCW Mobile and WeChat), since the tasks that the platform publicizes increase the data traffic, thereby benefiting the service providers. Therefore, the platform utility consists of three parts: the sensing performance, the revenue splits from the BS, and the rewards paid to the users. We give the platform utility as below:
\begin{align} \label{user_centric_noncooperative_utility}
{U_{nonco}} = \sum\limits_{i \in {\cal{N}}} {\Gamma_i} (\emph{\textbf{X}}) + \sum\limits_{i \in {\cal{N}},j \in {\cal{M}}} \gamma {{R_{i,j}}} (\emph{\textbf{X}}) - {\alpha_1}\sum\limits_{i \in {\cal{N}},j \in {\cal{M}}} {{Q_{i,j}} {x_{i,j}}}.
\end{align}
The utility of any user $j \in {\cal{M}}$ is then given by:
\begin{align} \label{user_centric_noncooperative_utility_user}
C_j^{nonco} = {\alpha_1} \sum\limits_{i \in \cal{N}} {{Q_{i,j}} {x_{i, j}}}  - \sum\limits_{i \in \cal{N}} {{R_{i,j}}} (\emph{\textbf{X}}).
\end{align}

Since the task matrix $\emph{\textbf{X}}$ and the subcarrier matrix $\emph{\textbf{S}}$ are decided separately by the users and the BS, we cannot formulate the problem as a joint optimization problem as the one in the centralized case. Note that the BS first allocates the subcarriers to the users, the users then decide which tasks to participate in based on the assigned subcarriers. We then formulate the non-cooperative approach as a Stackelberg game, which will be explained in detail in Section 3.


\subsection{Cooperative approach}%
In the traditional non-cooperative approach, each user makes independent decisions when choosing the sensing tasks and their payoff is only related to their contributions to the tasks. However, we may have a situation where too many users participate in one task $i$ due to its high input-output ratio, ${\varphi_i}/{\rho_i}$, such that task $i$'s performance reaches the upper bound $\varphi_i$, and the wireless subcarrier resources of some users are wasted.

To avoid the unequal distribution of resources, we propose the cooperative approach in which the users involved in the same task form a coalition, and the platform rewards the coalition instead of individual users, based on the task's performance instead of the individual contribution of each user. Therefore, the users will not participate in those tasks with enough users and saturated wireless channel resources when maximizing their utility. The waste of subcarrier resources can be effectively avoided in the cooperative approach.

We assume that the rewards for each task are proportional to the performance of this task, i.e., for any task $i \in \cal{N}$, the platform offers ${\alpha _2}{\Gamma_i}\left( \emph{\textbf{X}} \right)$ to the users completing this task, with ${\alpha_2}$ as a scaling factor describing the incentive intensity. The platform utility function is given by:
\begin{align} \label{user_centric_cooperative_utility}
{U_{co}} = \sum\limits_{i \in {\cal{N}}} {\Gamma_i} (\emph{\textbf{X}}) + \sum\limits_{i \in {\cal{N}},j \in {\cal{M}}} \gamma {{R_{i,j}}} (\emph{\textbf{X}}) - {\alpha_2}\sum\limits_{i \in {\cal{N}}} {\Gamma_i} (\emph{\textbf{X}}).
\end{align}
The utility of any user $j$ can be given by
\begin{align} \label{user_centric_cooperative_user}
C_j^{co} = \sum\limits_{i \in \cal{N}} {{p_{i,j}}}  - \sum\limits_{i \in \cal{N}} {{R_{i,j}}} (\emph{\textbf{X}}),
\end{align}
where ${{p_{i,j}}}$ represents the rewards that user $j$ gains from the platform for participating in task $i$, also known as user $j$'s payoff for partaking in task $i$. The specific form of ${{p_{i,j}}}$ is determined by the incentive mechanism, which will be discussed in Section \textbf{4}. Note that no matter which incentive mechanism is chosen, the rewards for any task $i \in {\cal{N}}$ are equal to the total payoff of the involved users, i.e., $\sum\nolimits_{j \in {\cal{M}}} {{p_{i,j}}}  = {\alpha _2}{\Gamma_i}(\emph{\textbf{X}})$.

In the cooperative approach, we assume that the subcarriers are allocated to the users through some simple methods, which will be described in Section \textbf{4}. Our goal is to design an incentive mechanism in which the users behave cooperatively to maximize their individual utility while the system can achieve a high utility. In Section \textbf{4}, we will discuss the cooperative approach in detail.

\section{Non-cooperative approach}

\subsection{Stackelberg game formulation}

As mentioned in Section \textbf{2.2}, the optimization problem in the non-cooperative approach can be divided into two subproblems: the task selection of the users and the subcarrier allocation of the BS. We model the non-cooperative approach as a Stackelberg game, in which there are two phases. In the first phase, the BS determines the allocation of the subcarriers to the users. In the second phase, each user selects multiple sensing tasks to participate in given the limit of their available resources. Therefore, the BS is the \emph{leader} and the users are the \emph{followers} in the Stackelberg game. The \emph{strategy of the BS} is the distribution of the subcarriers, i.e., the subcarrier matrix $\textbf{\emph{S}}$, while the \emph{strategy of user $j$} is the set of tasks that it participates in. The set of strategies of all the users can be expressed by the task matrix $\emph{\textbf{X}}$.


\subsubsection{Task selection of the users}
Suppose the subcarriers have been assigned to the users already, i.e., the subcarrier matrix $\emph{\textbf{S}}$ is given. Each user $j$ needs to choose which tasks to participate in, and to invest the feedback rate into these tasks. For any user $j$, it aims at maximizing its own utility $C_j^{nonco}$ with the limit of available data feedback rate, which can be formulated as a 0-1 knapsack problem. Given the subcarrier allocation, we assume that each user $j$ has a backpack with the capacity of $\sum\nolimits_{k \in {\cal{K}}} {{c_{k,j}}{s_{k,j}}}$. Each task $i \in \cal{N}$ is considered as an object with the volume of $r_i$ and the value of ${\alpha_1} {Q_{i,j}}  - \beta {r_i}$. Every object can only be picked and put into the backpack at most once. The goal of user $j$ is to maximize the total value of the backpack so that the sum of the volumes must be less than the knapsack's capacity. Thus for any user $j \in \cal{M}$, the task selection problem can be formulated as:
\begin{align} \label{0-1_pack_problem}
\begin{split}
& \mathop {\max }\limits_{\emph{\textbf{X}}} \sum\limits_{i \in {\cal{N}}} {\left( {{\alpha_1} {Q_{i,j}} - \beta {r_i}} \right){x_{i,j}}}, \quad {x_{i,j}} \in \left\{ {0,1} \right\} \\
& s.t.\sum\limits_{i \in {\cal{N}}} {{r_i}{x_{i,j}}}  \le \sum\limits_{k \in {\cal{K}}} {{c_{k,j}}{s_{k,j}}}.
\end{split}
\end{align}

The 0-1 knapsack problem has been proved to be an NP-hard one in which the closed form solution cannot be given \cite{14CLRS-2009}. We perform a linear relaxation method to obtain a closed-form sub-optimal solution which will be described in detail in Section 3.2.1.

\subsubsection{Subcarrier allocation of the BS}
The BS allocates the subcarriers without knowing how the users select the tasks. Note that the task matrix $\emph{\textbf{X}}$ can be estimated with the help of the platform under the condition of limited data feedback rate, i.e. $\sum\limits_{k \in {\cal{K}}} {{s_{k,j}}{c_{k,j}}}  \ge \sum\limits_{i \in {\cal{N}}} {{r_i}{x_{i,j}}}$. Since the interests of the BS and the platform are consistent with each other\footnote{The BS divides part of the charge of the feedback rate to the platform as revenue splits, so the incline of the platform utility also brings benefits for the BS.}, the platform can transmit the estimated X to the BS. The BS then maximizes the platform utility by determining the subcarrier matrix $\emph{\textbf{S}}$.
\begin{align} \label{non-cooperative_expect}
\begin{split}
& \mathop {\max }\limits_{\emph{\textbf{S}}} {U_{nonco}}\left( {E\left( \emph{\textbf{X}} \right)\left| {_{\sum\limits_{k \in {\cal{K}}} {{s_{k,j}}{c_{k,j}}}  \ge \sum\limits_{i \in {\cal{N}}} {{r_i}{x_{i,j}}} }} \right.} \right) \\
& s.t.  \sum\limits_{j \in {\cal{M}}} {{s_{k,j}} \le 1}.
\end{split}
\end{align}
We will discuss this problem in detail in Section 3.2.2.

\subsection{Design of the non-cooperative algorithm}
We first present the following theorem to state that there exists a Stackelberg equilibrium in the formulated problem.

\textbf{Definition 1:} A pair of strategies $\left( \emph{\textbf{X}}, \emph{\textbf{S}} \right)$ is a Stackelberg equilibrium if no unilateral deviation in strategy by the leader or the follower is profitable, i.e.,
\begin{equation} \label{stackel1}
\begin{split}
{U_{nonco}}\left( {\emph{\textbf{S}},\emph{\textbf{X}}} \right) & \ge {U_{nonco}}\left( {\emph{\textbf{S}},\emph{\textbf{X}}'} \right) \\
C_j^{nonco}\left( {\emph{\textbf{S}},{\emph{\textbf{x}}_j}} \right) & \ge C_j^{nonco}\left( {\emph{\textbf{S}},{\emph{\textbf{x}}_j}^\prime } \right).
\end{split}
\end{equation}

\textbf{Theorem 1:} There exists a Stackelberg equilibrium in the formulated Stackelberg game.
\begin{proof}
For a given subcarrier allocation scheme $\emph{\textbf{S}}_t$, there exists an optimal task selection solution for each user $i$, i.e., an optimal solution for the 0-1 knapsack problem of user $i$. The set of best response of the users is denoted as $\textbf{\emph{X}}^*$. Since there are finite subcarrier allocation schemes for the BS, there always exists a set of best response of the users i.e., $\textbf{\emph{X}}^*_t$, for each subcarrier allocation scheme $\emph{\textbf{S}}_t$. Suppose the BS can estimate the users' best response $\textbf{\emph{X}}^*_t$ according to their personal information. We can then find a subcarrier allocation scheme $\emph{\textbf{S}}_{op}$ such that the BS can obtain the highest utility by estimating the users' best response. The set of the users' best response can be obtained as $\textbf{\emph{X}}^*_{op}$. Therefore, no player tends to change its current strategy with the others' strategies unchanged. We then say $\left( \emph{\textbf{S}}_{op}, \textbf{\emph{X}}^*_{op} \right)$ is a Stackelberg equilibrium.
\end{proof}
\textbf{Remark 1:} The Stackelberg equilibrium cannot be reached within polynomial time.

Since there is no closed-form optimal solution for the 0-1 knapsack problem, the BS cannot precisely estimate the users' best response\footnote{When the BS estimates the users' best response to the subcarrier allocation, it needs a closed-form solution of the knapsack problem so as to present the subcarrier allocation problem in a mathematical form based on the estimated solution.}, and thus the exact equilibrium point of the Stackelberg game cannot be reached. Instead, we find a sub-optimal equilibrium by utilizing a non-cooperative algorithm in which the BS estimates the best response of the users via a linear relaxation method.

\subsubsection{Each follower's strategy}
We assume that each user selects the tasks by using the linear relaxation method~\cite{45D-1957}. Within the limit of feedback rate, a user selects its most preferred task first then the second preferred, the third preferred, and so on. To describe each user's preference on the tasks, we define a $N \times M$ preference matrix as $\emph{\textbf{T}} = {[{t_{i,j}}]_{N \times M}}$, where $t_{i,j}$ is the serial number of the $i$th-preferred task for user $j$. For example, ${t_{2,1}} = 3$ means that user 1's second preferred task in $\cal{N}$ is task 3. We assume that user $j$ prefers task $p$ to task $q$ if ${Q_{{t_{p,j}},j}}/{r_{{t_{p,j}}}} \ge {Q_{{t_{q,j}},j}}/{r_{{t_{q,j}}}}$, and thus we have: ${Q_{{t_{1,j}},j}}/{r_{{t_{1,j}}}} \ge {Q_{{t_{2,j}},j}}/{r_{{t_{2,j}}}} \ge  \cdots  \ge {Q_{{t_{N,j}},j}}/{r_{{t_{N,j}}}},\forall j \in \cal{M}$. Therefore, each user $j$'s strategy $\textbf{\emph{x}}_j$ can be presented as below:
\begin{align} \label{x_value_redefine}
{x_{i,j}}\left( {\bf{\emph{S}}} \right) = \left\{ {\begin{array}{*{20}{l}}
{1,\quad \sum\limits_{k \in \cal{K}} {{c_{k,j}}{s_{k,j}}}  \ge \sum\limits_{q = 1}^i {{r_{{t_{q,j}}}}} ,}\\
{0,\quad \mbox{otherwise}.}
\end{array}} \right.
\end{align}

\subsubsection{The leader's strategy}

The platform can obtain the preference matrix by either the users' reports to it or machine learning based analysis on the users' behaviours. The platform then report the preference matrix $\emph{\textbf{T}}$ to the BS, and the BS predicts that user $j$ will participate in its $i$th-preferred task $t_{i,j}$ only when the total feedback rate of user $j$'s first $i$ preferred tasks is smaller than user $j$'s available feedback rate $\sum\limits_{k \in \cal{K}} {{c_{k,j}}{s_{k,j}}}$. With the estimated $\emph{\textbf{X}}$, the maximization problem in $\left( \ref{non-cooperative_expect} \right)$ can be mathematically formulated as a 0-1 INLP problem:
\begin{align} \label{noncooperative_utility_maximazition}
\begin{split}
& \mathop {\max }\limits_{\emph{\textbf{S}}} {U_{nonco}}\left( {\emph{\textbf{X}}(\emph{\textbf{S}})} \right), \quad {s_{k,j}} \in \{ 0,1\} \\
& s.t.\sum\limits_{j \in \cal{M}} {{s_{k,j}}}  \le 1,
\end{split}
\end{align}
where $\emph{\textbf{X}}(\emph{S})$ is given by $\left( \ref{x_value_redefine} \right)$. The formulated 0-1 INLP problem is NP-hard and the exact solution takes exponential complexity \cite{12RR-2009}\cite{30BSS-2013}. When we solve the 0-1 INLP problem formulated in (18), a local optimal solution can be found by splitting the INLP problem into multiple spherical LP problems, which can then be solved utilizing the branch and bound method. As the scale of the network increases, the computational complexity of this algorithm can be significantly high due to the growing number of variables and constraints\footnote{The 0-1 INLP problem formulated in the centralized case can be solved utilizing similar methods, but the complexity of the centralized case is even higher than that of the non-cooperative approach, since there are more variables to be settled in the centralized case with the same network parameters.}.

\subsubsection{Description of the non-cooperative algorithm}

We now describe the overall algorithm for the non-cooperative approach. In phase 1, before the BS allocates the subcarriers to the users, it will obtain a preference matrix from the platform, and it assumes that the users will select the tasks according to the preference matrix. Then the BS tries to maximize the expected platform utility by allocating the subcarriers to the users. The BS formulates the problem as a 0-1 INLP problem and obtains the subcarrier matrix by solving it. In phase 2, given the assigned subcarrier resources, each user maximizes its own utility by selecting the tasks to participate in. The overall non-cooperative optimization algorithm is presented in Table \ref{AlgorithmFormation1}.

Note that the optimal solution of the non-cooperative approach is to solve this problem as a joint optimization one, which is also a NP-hard problem. However, this is not practical since the BS and the users make their decisions separately. The algorithm proposed here is a step-by-step approach, in which the independence of the BS and the users are considered, and each of the two steps in Table \ref{AlgorithmFormation1} can be solved using existing literature. Therefore, we don't guarantee the optimality of the solution, but it gives a traditional method of designing an incentive mechanism in which the limit of wireless channel resources is considered. The non-cooperative approach can serve as a comparison to the cooperative approach, which will be presented in detail in Section \textbf{5}.

The computational complexity of the non-cooperative algorithm mainly lies in solving the 0-1 INLP problem in $\left( \ref{noncooperative_utility_maximazition} \right)$. We use an optimization software LINGO\cite{LINGO} to solve this problem, and record the number of iterations to evaluate the complexity of this algorithm. The comparison result with the cooperative approach will be presented in Table~\ref{complexity} in Section \textbf{5}.

\begin{table}[!t]
\renewcommand{\arraystretch}{1.3}
\caption{Non-cooperative Optimization Algorithm}
\label{AlgorithmFormation1}
\centering
\begin{tabular}{p{150mm}}

\hline

\textbf{Phase 1: Subcarrier allocation to the users:}

\begin{enumerate}
    \item The platform obtains the preference matrix $\emph{\textbf{T}}$ by either behaviour analysis based machine learning or the users' reports to the platform.
    \item The platform transmits the preference matrix $\emph{\textbf{T}}$ to the BS.
    \item With the preference matrix $\emph{\textbf{T}}$, the BS redefine the task matrix $\emph{\textbf{X}}$ as presented in $\left( \ref{x_value_redefine} \right)$, and solves the maximization problem formulated in $\left( \ref{noncooperative_utility_maximazition} \right)$.
    \item The BS assigns the set of subcarriers $\cal{K}$ according to the solution got from \textbf{Step 1}-1.
\end{enumerate}

\textbf{Phase 2: Task selection by the users:}

\begin{enumerate}
    \item Each user formulates its task selection problem as a 0-1 knapsack problem presented in $\left( \ref{0-1_pack_problem} \right)$, and solves it using the linear relaxation method.
    \item Each user $j$ distributes its feedback rate to the tasks according to the solution got from \textbf{Step 2}-1.
\end{enumerate}
\\

\hline

\end{tabular}
\end{table}
\vspace{-0.1cm}
\section{Cooperative approach using overlapping coalition formation games}%
In this section, we assume that the subcarriers are allocated by the BS through either the random allocation or the priority-based allocation, which will be explained in detail in Section \textbf{4.2.1}. The users participate in more than one task in most cases so as to maximize their utility. Specifically, they may tend to cooperate with each other while allocating their resources in order to avoid the waste of their resources. Based on this, we focus on the cooperative behaviors of the users, and introduce the~OCF-game in which the users form overlapping coalitions to participate in the tasks. We assume that the members of each coalition contribute the feedback rate to this coalition, and obtain rewards from the platform according to the performance of the corresponding task. Based on the OCF-game model, an overlapping coalition formation algorithm (OCF-algorithm) is then proposed for the users, which converges to a stable overlapping coalition structure (OCS).

\subsection{Overlapping Coalition Formation Game Formulation}

We begin by describing some notations and the model of OCF games. When presenting the definitions, we make some modification while mostly following those in~\cite{23CEW-2012},~\cite{33ZE-2011},~\cite{35SHDHB-2009}, so as to fit our scenario better.

In the cooperative approach, any user $j \in \cal{M}$ is seen as a player, and the resources of each user are the allocated feedback rate $\sum\nolimits_{k \in \cal{K}} {{c_{k,j}}{s_{k,j}}}$. User $j$'s \emph{strategy} is denoted as ${\emph{\textbf{b}}_j} = (b_j^1, \cdots ,b_j^N)$, where $b_j^i = 1$ denotes that user $j$ participates in task $i$ and invests $r_i$ bits of resources into this task. The set of all the strategies of user $j$ is denoted as ${\emph{\textbf{B}} _j}$.

With the notion of $b_j^i$, we then define the \emph{coalition} for task $i \in \cal{N}$ as ${\emph{\textbf{b}}^i} = (b_1^i, \cdots ,b_M^i)$, where $b_j^i = 1$ also represents that user $j$ is a member of coalition $i$. The support of coalition $\emph{\textbf{b}}^i$, denoted by $supp(\emph{\textbf{b}}^i)$, is defined as $supp({\emph{\textbf{b}}^i}) = \{ j \in {\cal M}|b_j^i = 1\}$, which represents the set of coalition members. Note that we allow the support of a coalition to be empty if there is no user participating in the corresponding task. As we have mentioned in Section \textbf{2}, once user $j$ joins any coalition $\emph{\textbf{b}}^i$, it needs to invest at least ${r_i}$ bits of resources into this coalition. Thus, to avoid individual resource waste, the users may only want to divide their resources in a discrete manner when investing. Note that the discrete manner of resource allocation guarantees the integral transmission of the sensed data, and avoids developing a vague bound on the number of potential coalition structures.

Now we give the value of a coalition $\emph{\textbf{b}}^i$ by a characteristic function $v$ : ${\left[ {0,1} \right]^M} \to {{\mathbb{R}}_ + }$. Based on the performance of a task that we have explained in $\left( \ref{onetask_performance} \right)$, the value of coalition $\emph{\textbf{b}}^i$ is then defined as:
\begin{equation} \label{Value}
v({\emph{\textbf{b}}^i}) = {\alpha _2}{\Gamma_i} =
\begin{cases}
{\frac{{\varphi_i}}{{\rho_i}}\sum\limits_{j \in \cal{M}} {{Q_{i,j}}} b_j^i}  , &~{\sum\limits_{j \in \cal{M}} {{Q_{i,j}}b_j^i}  \le \rho_i,} \\
\varphi_i,  &~ \mbox{otherwise},
\end{cases}
\end{equation}
where ${\alpha _2}{\Gamma_i}$ is the total rewards that the platform gives to the users participating in task $i$, and ${Q_{i,j}}b_j^i$ is the contribution that user $j$ makes to coalition $\emph{\textbf{b}}^i$. The characteristic form of the value function $\left( \ref{Value} \right)$ implies that the value of a coalition is completely decided by the members of the coalition. To be specific, the value of the coalition depends on the coalition members' contribution to it.

With all the concepts mentioned above, the proposed OCF-game is then defined as below.

\textbf{Definition 2:} An \emph{OCF-game} $G = \left( {{\mathcal{M}},v} \right)$ is defined by a set of users ${\cal{M}} = \{ 1, \cdots ,M\}$ and a value function $v:{\{ 0,1\} ^M} \to {{\mathbb{R}}^ + }$ where $v({0^M}) = 0$. The characteristic form of the value function is given in $\left( \ref{Value} \right)$.

In an \emph{OCF-game} $G = \left( {{\mathcal{M}},v} \right)$, an \emph{overlapping coalition structure} (OCS) over $M$ is a $M \times N$ matrix $\Theta = \left( {{\emph{\textbf{b}}_1}, \cdots ,{\emph{\textbf{b}}_N}} \right)$, where $N$ is the number of coalitions. Since we have $N$ tasks corresponding to $N$ coalitions in the system, the size of an OCS is fixed to be $N$. It is also required that $\sum\limits_{i = 1}^N {b_j^i{r_i}}  \le \sum\nolimits_{k \in {\cal{K}}} {{c_{k,j}}{s_{k,j}}}$, which guarantees that $\Theta$ is a valid division of the users' resources.


The value of coalition $\emph{\textbf{b}}^i$ is also the payoff that needs to be divided among the users who contribute to coalition $\emph{\textbf{b}}^i$, i.e., the members of $supp(\emph{\textbf{b}}^i)$. Here, we define the \emph{payoff distribution} as a finite list of vectors $\emph{\textbf{P}} = \{\emph{\textbf{p}}^1,\ldots,\emph{\textbf{p}}^N\}$, and $\emph{\textbf{p}}^i \in \mathbb{R}^M$ is the payoff vector for the members in coalition $\emph{\textbf{b}}^i$ that satisfies $\sum\limits_{j = 1}^M {{p_{i,j}}}  = v\left( {{\emph{\textbf{b}}^i}} \right)$.

We assume that the value of a coalition is assigned to and only to its coalition members due to their participation levels, and the payoff of any coalition member is unaffected by those outsiders. For any user $j \in \cal{M}$, the payoff obtained from coalition $\emph{\textbf{b}}^i$ is the rewards that user $j$ gets from the platform for participating in task $i$, which can be mathematically given by
\begin{align} \label{Payoff}
p_{i,j} = {\phi _j}({\emph{\textbf{b}}^i}) = \left\{ {\begin{array}{*{20}{l}}
v({\emph{\textbf{b}}^i})\frac{{{Q_{i,j}}}}{{\sum\limits_{q \in supp({\emph{\textbf{b}}^i})} {{Q_{i,q}}} }}, &{j \in supp({\emph{\textbf{b}}^i}),}\\
{0,}&{j \notin supp({\emph{\textbf{b}}^i}).}
\end{array}} \right.
\end{align}
With the utility function $\left( \ref{user_centric_cooperative_user} \right)$ and the payoff equation $\left( \ref{Payoff} \right)$, we can then obtain the specific form of a user's utility, which is omitted here.

Note that the charge of data rate and the value function of a coalition are two important factors of governing each coalition's size. As the size of a coalition increases, the value of the coalition grows and will stop increasing at some point where each user's payoff begins to decrease. Thus, no additional user would like to join the coalition at which point it will not gain any positive utility.


\subsection{Design of Overlapping Coalition Formation Algorithm}%

In our proposed OCF-game, the direct motivation of user $j \in \mathcal{M}$ is to increase its utility $C_j^{co}$. As mentioned in Section \textbf{4.1}, user $j$ may not gain positive utility from joining a coalition, since in some cases the payoff that user $j$ obtains from this coalition is less than the charge of feedback rate. Therefore, user $j$ needs to consider carefully when joining a new coalition. Besides, the other members of this coalition have right to decide whether to accept this user according to their own utility. Likewise, a user also needs to consider quitting a coalition if it cannot obtain any positive utility from this coalition.

For the convenience of discussion, we classify the fundamental operations of a user into three categories: \emph{transfer operation}, quitting a coalition and joining a new coalition. Each kind of operations will lead to a change in OCS. We will then introduce the concepts of these three operations and give the corresponding execution conditions in our proposed OCF-game.

\textbf{Definition 3:} For any user $j \in \mathcal{M}$ in the proposed OCF-game with the current OCS $\Theta = \{\emph{\textbf{b}}^1,\ldots,\emph{\textbf{b}}^N\}$, a \emph{transfer operation} from coalition $\emph{\textbf{b}}^p \in \Theta$ to coalition $\emph{\textbf{b}}^q \in \Theta$, denoted by $T_j(\emph{\textbf{b}}^p,\emph{\textbf{b}}^q)$, is to withdraw all of user $j$'s invested resources, i.e., $r_p$, from coalition $\emph{\textbf{b}}^p$ and invest the required amount of resources, i.e., $r_q$, into coalition $\emph{\textbf{b}}^q$.

After user $j$ withdraws its resources from coalition $\emph{\textbf{b}}^p$, it quits this coalition. Coalition $\emph{\textbf{b}}^p$ then becomes $\emph{\textbf{b}}^{p-}$, which satisfies $b_{_j}^{p - } = 0$, i.e., user $j$ is not a member of coalition $\emph{\textbf{b}}^{p-}$. Similarly, after user $j$ transfers its resources to coalition $\emph{\textbf{b}}^q$, it joins this coalition as a new member. Coalition $\emph{\textbf{b}}^q$ then becomes $\emph{\textbf{b}}^{q+}$ satisfying $b_{_j}^{q + } = 1$. The new OCS is expressed as $\Theta' = \Theta \backslash \{\emph{\textbf{b}}^p, \emph{\textbf{b}}^q\} \cup \{ \emph{\textbf{b}}^{p-}, \emph{\textbf{b}}^{q+} \}$. Note that the above notions $\emph{\textbf{b}}^{p-}$ and $\emph{\textbf{b}}^{q+}$ only take effect within a transfer operation to describe the change of the coalitions.

When we further judge whether the \emph{transfer operation} $T_j(\emph{\textbf{b}}^p,\emph{\textbf{b}}^q)$ is feasible for user $j \in \cal{M}$, a series of conditions should be satisfied first. Considering the limited feedback rate of user $j$, we have:
\begin{align} \label{realizable}
\sum\limits_{i \in \mathcal{N}\backslash \left\{ {p, q} \right\}} {{r_i}b_j^i}  - {r_p} + {r_q} \le \sum\limits_{k \in \mathcal{K}} {{c_{k,j}}{s_{k,j}}}.
\end{align}

If the utility that user $j$ gets from joining coalition $\emph{\textbf{b}}^q$ is positive and larger than that from joining coalition $p$, the \emph{transfer operation } $T_j(\emph{\textbf{b}}^p,\emph{\textbf{b}}^q)$ is \emph{profitable} for user $j$ itself, i.e.,
\begin{align} \label{profitable}
{\phi _j}\left( {{{\bf{b}}^{q + }}} \right) - \beta {r_q} > \max \left\{ {0,{\phi _j}\left( {{{\bf{b}}^{p - }}} \right) - \beta {r_p}} \right\}.
\end{align}

However, even if the \emph{transfer operation} ${T_j}\left( {{{\textbf{\emph{b}}}^p},{{\emph{\textbf{b}}}^q}} \right)$ is profitable for user $j$, it is not necessary that this \emph{transfer operation} is feasible. The payoff of other users in coalition $\emph{\textbf{b}}^q$ should be considered as well due to the characteristics of the coalitions. Normally, when a user $j$ tends to invest its resources to coalition $\emph{\textbf{b}}^q$, the members of coalition $\emph{\textbf{b}}^q$ has right to decide whether to accept this user. If the utility of other users in coalition $\emph{\textbf{b}}^q$ is affected when user $j$ joins this coalition, then these users are not willing to let user $j$ in. Note that user $j$ removes all its resources from coalition $\emph{\textbf{b}}^p$ when executing the \emph{transfer operation}, so it has no stake in what other members in coalition $\emph{\textbf{b}}^p$ will react. Therefore, before user $j$ makes a \emph{transfer operation} ${T_j}({\emph{\textbf{b}}^p},{\emph{\textbf{b}}^q})$, it needs to obtain the information from other members of coalition $\emph{\textbf{b}}^q$ in order to judge whether this \emph{transfer operation} is \emph{permitted}. We then formally define a \emph{transfer operation} ${T_j}({\emph{\textbf{b}}^p},{\emph{\textbf{b}}^q})$ as \emph{permitted} by all of the other members in coalition $\emph{\textbf{b}}^q$ if it satisfies:
\begin{align} \label{permitted}
{\phi _k}({\emph{\textbf{b}}^{q + }}) \ge {\phi _k}({\textsl{\textbf{b}}^q}), \forall k \in \{ supp({\emph{\textbf{b}}^q})|k \ne j\}.
\end{align}


With all the conditions mentioned above, the feasibility of a \emph{transfer operation} can be defined as follows:

\textbf{Definition 4:} In the proposed OCF-game with the current OCS $\Theta = \{\emph{\textbf{b}}^1,\ldots,\emph{\textbf{b}}^N\}$, the \emph{transfer operation } ${T_j}({\emph{\textbf{b}}^p},{\emph{\textbf{b}}^q})$ is \emph{feasible} and can be executed if it satisfies $\left( \ref{realizable} \right)$, $\left( \ref{profitable} \right)$, $\left( \ref{permitted} \right)$.

However, the \emph{transfer operation} is not the only action that the users consider while shifting their resource allocation. A user $j$ can choose to just quit a coalition $\emph{\textbf{b}}^p$, and not to reinvest the resources into any other coalitions. We assume that once user $j$ quits coalition $\emph{\textbf{b}}^p$, it withdraws all its resources invested into this coalition, i.e., $b_j^p = 0$, and user $j$ gets no payoff from coalition $\emph{\textbf{b}}^p$. User $j$ considers quitting coalition $\emph{\textbf{b}}^p$ when it cannot obtain positive utility from this coalition or any potential \emph{transfer operations}. Therefore, it is \emph{feasible} for any user $j$ to quit coalition $\emph{\textbf{b}}^p$ if:
\begin{align} \label{quit_coalition}
\begin{cases}
{\phi _j}({{\emph{\textbf{b}}}^p}) - \beta {r_p} \le 0, \\
{T_j}\left( {{{\emph{\textbf{b}}}^p},{{\emph{\textbf{b}}}^q}} \right) \text{is not profitable for user $j$},\forall q \in \{ {\cal{M}}|b_j^q = 0\}.
\end{cases}
\end{align}

Likewise, when shifting its resource allocation, user $j$ can also choose to just join a new coalition $\emph{\textbf{b}}^q$, and invest its resources into this coalition. This consideration is reasonable since it is possible that after user $j$ executes a transfer operation ${T_j}({\emph{\textbf{b}}^s},{\emph{\textbf{b}}^t})$, it may have extra resources uninvested, so user $j$ may consider joining a new coalition $\emph{\textbf{b}}^q$. For its own sake, user $j$ only considers joining a new coalition $\emph{\textbf{b}}^q$ when it has enough resources to invest and can get positive utility from coalition $\emph{\textbf{b}}^q$. Besides, the utility of other members in coalition $\emph{\textbf{b}}^q$ should be considered as well. Therefore, it is \emph{feasible} for any user $j$ to join a new coalition $\emph{\textbf{b}}^q$ if:
\begin{align} \label{join_coalition}
\begin{cases}
{\phi _j}({{\emph{\textbf{b}}}^q}) - \beta {r_q} > 0, \\
\sum\limits_{i \in \mathcal{N}\backslash \left\{ q \right\}} {b_i^j{r_i}}  + {r_q} \le \sum\limits_{k \in {\cal{K}}} {{c_{k,j}}{s_{k,j}}}, \\
{\phi _k}({{\emph{\textbf{b}}}^{q + }}) \ge {\phi _k}({{\emph{\textbf{b}}}^q}),\forall k \in \{ supp({{\emph{\textbf{b}}}^q})|k \ne j\}.
\end{cases}
\end{align}

\subsubsection{Algorithm Description}
We now describe the proposed OCF-algorithm, which is designed for solving the resource allocation problem of the users in a distributed way. The algorithm consists of two phases: the subcarrier allocation phase and the coalition formation phase as shown in Table 2.

In the first phase, the BS assigns the subcarriers to the users through the random allocation or the priority-based allocation. In the random allocation method (Step 1-1), each subcarrier is randomly assigned to a user in the system. In the priority-based allocation (Step 1-2), the platform first collects the users' location information with the help of the location-based service. It then evaluates the potential contributions that each user $j$ can make to all the tasks according to $R{K_j} = \sum\nolimits_{i \in \cal{N}} {{Q_{i,j}}} /{r_i}$. Each user's priority is determined by the BS based on the decreasing order of $\{ R{K_j}\} _{j = 1}^M$. In each round of subcarrier allocation, the BS allocates the unassigned subcarriers to the set of users one by one according to the priority list, i.e., user $p$ is allocated its currently most preferred subcarrier among the unassigned subcarriers ahead of user $q$ if $R{K_p} > R{K_q}$. Each user is allocated one subcarrier in each round, and the iteration stops until all the subcarriers have been allocated.

In the second phase, for a given value of $\alpha_2$, each user initially chooses a set of coalitions to join utilizing a linear relaxation algorithm, i.e., the same algorithm that the platform uses to predict the users' behaviors in Section \textbf{3.2} (Step 2-1 and Step 2-2). Based on the initial set of coalitions, multiple iterations are performed by the users until a stable OCS is formed. We now provide a brief description of the algorithm during one iteration, i.e., from Step 2-3 to Step 2-8. In Step 2-3 and 2-4, each user $j$ first records its current \emph{strategy} as ${\textbf{\emph{b}}_j}$, then it considers whether to quit coalition $i$ if it can't get positive utility from this coalition. User $j$ can only quit coalition ${\textbf{\emph{b}}^i}$ when its quitting does not damage other members' profits. In Step 2-5, each user $j$ keeps trying the \emph{transfer operation} between corresponding coalitions, until there is no \emph{feasible} \emph{transfer operations} for user $j$. After executing the \emph{transfer operations}, each user $j$ will check whether it is \emph{feasible} join any new coalitions. From Step 2-6 to 2-8, each user $j$ records its current \emph{strategy} as ${\textbf{\emph{b}}_j}^\prime $ after all the operations performed above. The iterations stop if each user's \emph{strategy} before and after shifting the resources remains the same. We say that the users converge to a stable OCS and they stop shifting their resources.

It should be noted that the resource allocation problem we solve here is an NP-hard problem \cite{40DJ-2006}, which means that the optimal solution can't be achieved within polynomial time. So it is normal that the final OCS is not guaranteed to be optimal. This can be explained in a straightforward way as below. In the network, the most important thing for the users is to maximize their own profits gained from the coalitions, so in most cases they are not fully cooperative. Therefore, the final OCS is not always socially optimal.

\begin{table}[!t]
\renewcommand\arraystretch{1.3}
\caption{Overlapping Coalition Formation Algorithm for Smartphone Sensing}
\label{distributedSS}
\centering
\begin{tabular}{p{150mm}}
\hline
\vspace{-0.1cm}
\textbf{Step 1: Subcarrier Allocation Phase}

The BS allocates the subcarriers to the users through either of the following methods:
\begin{enumerate}
\vspace{-0.1cm}
    \item random allocation: for all $k \in \cal{K}$
    \begin{enumerate}
        \item ${j^*} = {\mbox{random}}(\{ j,\forall j \in {\cal{M}}\} )$.
        \vspace{-0.1cm}
        \item allocate subcarrier $k$ to user ${j^*}$.
    \end{enumerate}
    \vspace{-0.1cm}
    \item priority-based allocation:
    \begin{enumerate}
        \item Define $\{ R{K_j}\} _{j = 1}^M$, $R{K_j} \in {\mathbb{R}}$, $\{ {\chi _j}\} _{j = 1}^M$, ${\chi _j} \in \{ 1,M\}$, and $\{ {\zeta _j}\} _{j = 1}^M$, ${\zeta _j} \in \{ 1,K\}$.
            \vspace{-0.1cm}
        \item Set $R{K_j} = \sum\nolimits_{i \in \cal{N}} {{Q_{i,j}}} /{r_i}$.
        \vspace{-0.1cm}
        \item Sort $R{K_j}$ from largest to smallest, then record the corresponding index $j$ in $\chi$ as ${\chi _j}$.
        \vspace{-0.1cm}
        \item For all ${\chi _j} \in \left\{ {1,M} \right\}$,
        \begin{enumerate}
        \vspace{-0.1cm}
            \item ${\zeta _{\chi _j}} = \arg \mathop {\max }\limits_{k \in {\cal{K}}} \{ {c_{k,{\chi _j}}}\}$.
            \vspace{-0.1cm}
            \item Assign subcarrier ${\zeta _{\chi _j}}$ to user $j$, and then remove subcarrier ${\zeta _{\chi _j}}$ from $\cal{K}$.
        \end{enumerate}
    \end{enumerate}
\end{enumerate}
\textbf{Step 2: Coalition Formation Phase}
\begin{enumerate}
    \item Obtain task priority vector ${\emph{\textbf{t}}_j} = ({t_{1,j}},{t_{2,j}}, \cdots ,{t_{N,j}})$ for all $j \in {\cal{M}}$.
    \vspace{-0.1cm}
    \item For all $i \in {\cal{N}}$ and all $j \in {\cal{M}}$,
    \begin{enumerate}
    \vspace{-0.1cm}
        \item If ${\sum\nolimits_{k \in \cal{K}} {{c_{k,j}}{s_{k,j}}}  \ge \sum\limits_{q = 1}^i {{r_{{t_{q,j}}}}} }$, ${b_j^i} = 1$.
        \vspace{-0.1cm}
        \item Else ${b_j^i} = 0$.
    \end{enumerate}
    \item Each user $j \in {\cal{M}}$ records its \emph{strategy} ${\textbf{\emph{g}}_j}$.
    \vspace{-0.1cm}
    \item For all $i \in {\cal{N}}$ and all $j \in {\cal{M}}$,
    \begin{enumerate}
    \vspace{-0.1cm}
        \item If ${p_{i,j}} - {R_{i,j}} < 0$, judge whether there exists a \emph{profitable} \emph{transfer operation } ${T_j}({{\emph{\textbf{b}}}^i},{{\emph{\textbf{b}}}^k})$ for user $j$.
        \vspace{-0.1cm}
        \begin{enumerate}
        \item If so, execute ${T_j}({{\emph{\textbf{b}}}^i},{{\emph{\textbf{b}}}^k})$.
        \item Else, if ${\phi _k}({{\emph{\textbf{b}}}^{i - }}) \ge {\phi _k}({{\emph{\textbf{b}}}^i}),\forall k \in \{ supp({{\emph{\textbf{b}}}^i})|k \ne j\}$ stands, user $j$ quits coalition $i$.
        \end{enumerate}
        \vspace{-0.1cm}
    \end{enumerate}
    \vspace{-0.1cm}
    \item For user $j = 1$ to $M$
        \begin{enumerate}
        \vspace{-0.1cm}
            \item Let $R \subseteq \Theta$ be the set of coalitions that contains user $j$.
            \vspace{-0.1cm}
            \item Select two coalitions $\emph{\textbf{b}}^{p,{\ell}}$ and $\emph{\textbf{b}}^{q,{\ell}}$ in $R$, the pair never shows; otherwise go to \textbf{Step 2}-5-d.
                \vspace{-0.1cm}
            \item If ${T_j}({\emph{\textbf{b}}^{p,{\ell}}},{\emph{\textbf{b}}^{q,{\ell}}})$ is \emph{feasible}, execute this operation and go to \textbf{Step 2}-5-b, else go back to \textbf{Step 2}-5-b.
            \vspace{-0.1cm}
            \item For all $i \in {\cal{N}}$, check if it is \emph{feasible} for user $j$ to join coalition $i$. If so, user $j$ joins coalition $i$.
            \item Turn to another user and go to \textbf{Step 2}-5-a.
        \end{enumerate}
        \vspace{-0.1cm}
    \item Set ${\eta} = 0$. For user $j = 1$ to $M$,
        \begin{enumerate}
        \vspace{-0.1cm}
        \item Each user $j$ records its \emph{strategy} as ${\textbf{\emph{g}}_j}^\prime $.
        \vspace{-0.1cm}
        \item If ${\textbf{\emph{g}}_j}$ is identical with ${\textbf{\emph{g}}_j}^\prime $, set ${\eta} = {\eta} + 1$ and ${\textbf{\emph{g}}_j} = {\textbf{\emph{g}}_j}^\prime $.
        \vspace{-0.1cm}
        \item Else set ${\textbf{\emph{g}}_j} = {\textbf{\emph{g}}_j}^\prime $.
        \end{enumerate}
        \vspace{-0.1cm}
    \item If ${\eta} = M$, go to \textbf{Step 2}-8.
    \item Else go to \textbf{Step 2}-3.
    \item Finalize the coalition structure and then go to \textbf{Step 3}.
\end{enumerate}
\textbf{Step 3: End of the algorithm.}\\
\hline
\end{tabular}
\end{table}

\subsection{Analysis of the OCF Algorithm}
\subsubsection{Stability and convergence}
Different from the traditional coalition formation games, the OCF-game is still an ongoing topic and there exist no general stability concepts \cite{33ZE-2011} \cite{34ZCE-2012}. In \cite{1CEMJ-2010}, the concept of \emph{core} is introduced to describe the stability of the OCF-games, however, this concept is not suitable for our proposed OCF-game model. First, the users' resources are limited in our scenario, while the users' resources are infinite in \cite{1CEMJ-2010}. Second, we consider the cost of the users for using the resources (i.e., the charge of data feedback rate) in our scenario, and thus, the users' behaviors are modeled slightly differently. Therefore, we incorporate the concept of stability in the traditional cooperative games \cite{35SHDHB-2009}\cite{36KA-2009} and extend it in our proposed OCF-algorithm by defining the concept of~\emph{T-stable} OCS as below.

\textbf{Definition 4:} In the proposed OCF-game with the OCS $\Theta = \{\emph{\textbf{b}}^1,\cdots,\emph{\textbf{b}}^N\}$, for any user $j \in {\cal{M}}$, if it cannot make any \emph{feasible} \emph{transfer operations}, or quit any coalitions, or join any new coalitions in its strategy ${\emph{\textbf{b}}_j} \in {\emph{\textbf{B}}_j}$ given the strategies of the other users $\left\{ {{\emph{\textbf{b}}_1}, \cdots ,{\emph{\textbf{b}}_M}} \right\}\backslash {\emph{\textbf{b}}_j}$, then we say the OCS $\Theta$ here is \emph{T-stable}.

In other words, a \emph{T-stable} OCS in our proposed OCF-game corresponds to an equilibrium state in which no user has the incentive to shift its resource allocation from the already formed coalitions. Therefore all the users stick to their current strategies and no change happens in the \emph{T-stable} OCS.

\textbf{Lemma 1}: If the OCF-algorithm can converge to a final OCS ${\Theta ^*}$, then we say this OCS is \emph{T-stable}.

\begin{proof}
According to our proposed OCF-algorithm in Table \ref{distributedSS}, when proposed OCF-algorithm converges to a terminal OCS ${\Theta ^*}$, each user $j$'s \emph{strategy} satisfies: (i) there is no \emph{feasible} \emph{transfer operations} for user $j$; (ii) there is no coalition that is \emph{feasible} for user $j$ to quit. Thus, the strategy that user $j \in \cal{M}$ takes must be the best strategy ${\emph{\textbf{b}}_j}^*$ for user $j$ in current situation, i.e., ${C_{j\_co}}({\Theta _{{\emph{\textbf{b}}_j}^*, {\emph{\textbf{b}}_{ - j}}}}) > {C_{j\_co}}({\Theta _{{\emph{\textbf{b}}_j}, {\emph{\textbf{b}}_{ - j}}}})$, $\forall {\emph{\textbf{b}}_j} \in {\textbf{B}_j}$. Therefore, there is no user $j \in \cal{M}$ that can improve its utility by a unilateral change in its strategy ${\emph{\textbf{b}}_j} \in {\emph{\textbf{B}}_j}$. Hence, the terminal OCS ${\Theta ^*}$ is \emph{T-stable}.
\end{proof}

\textbf{Theorem 2}: Starting from the initial OCS, our proposed OCF-algorithm based on the users' shift in their resource allocation converges to a \emph{T-stable} OCS after limited iterations.

\begin{proof}
After the initiation of the users' strategies, the initial OCS is ${\Theta _0}$. Based on the iterations in which the users try to maximize their utility, we can express the change of OCS as follows:
\begin{align} \label{iteration}
{\Theta _0} \to {\Theta _1} \to {\Theta _2} \to  \cdots.
\end{align}

After iteration ${\ell}$, the OCS changes from ${\Theta_{\ell - 1}}$ to ${\Theta_{\ell}}$, and the \emph{strategy} of any user $j$ changes from $\emph{\textbf{b}}_j ({{\ell} - 1})$ to $\emph{\textbf{b}}_j({\ell})$, which satisfies $C_j^{co,{\ell} - 1} < C_j^{co,{\ell}}$. Since the \emph{transfer operation} $T_j(\emph{\textbf{b}}^p,\emph{\textbf{b}}^q)$ is guaranteed to be a \emph{feasible} one, the utility of both user $j$ and other members of coalition $\emph{\textbf{b}}^q$ will not be affected due to user $j$'s shift in resource allocation. As for the members of coalition $\emph{\textbf{b}}^p$ which user $j$ quits, we prove that their utility will not be affected by $T_j(\emph{\textbf{b}}^p,\emph{\textbf{b}}^q)$ as below. According to equation $\left( \ref{Payoff} \right)$, the payoff of any user $k$ obtained from coalition $\emph{\textbf{b}}^p$ before user $j$ quits can be expressed as:
\begin{equation} \label{proof_otheruser1}
{p_{p,k}} = {\phi _k}({\emph{\textbf{b}}^p}) = v({\emph{\textbf{b}}^p})\frac{{{Q_{p,k}}}}{{\sum\limits_{j \in supp({\emph{\textbf{b}}^p})} {{Q_{p,j}}} }},
\end{equation}
and the payoff that any user $k$ obtains from coalition $\emph{\textbf{b}}^p$ after user $j$ quits is:
\begin{equation} \label{proof_otheruser2}
{p_{p,k}}^\prime  = {\phi _k}({\emph{\textbf{b}}^{p - }}) = v({\emph{\textbf{b}}^{p - }})\frac{{{Q_{p,k}}}}{{\sum\limits_{j \in supp({\emph{\textbf{b}}^{p - }})} {{Q_{p,j}}} }}.
\end{equation}
Given the value function of coalition $\emph{\textbf{b}}^p$ in $\left( \ref{Value} \right)$, the size relationship of $v({\emph{\textbf{b}}^p})$ and $v({\emph{\textbf{b}}^{p - }})$ can be divided into three conditions:

(i) $v({{\textbf{\emph{b}}}^p}) < \varphi_p$,
\begin{equation}\label{deduction1}
\begin{split}
& \frac{{v({{\textbf{\emph{b}}}^p})}}{{\sum\limits_{j \in supp({{\textbf{\emph{b}}}^p})} {{Q_{p,j}}} }} = \frac{{v({{\textbf{\emph{b}}}^{p - }})}}{{\sum\limits_{j \in supp({{\textbf{b}}^{p - }})} {{Q_{p,j}}} }} = \frac{{\varphi_p}}{{\rho_p}} \\
\Rightarrow \quad & {p_{p,k}} = {p_{p,k}}^\prime.
\end{split}
\end{equation}

(ii) $v({{\textbf{\emph{b}}}^p}) = \varphi_p$, and $v({\emph{\textbf{b}}^{p - }}) < \varphi_p$,
\begin{equation}\label{deduction2}
\begin{split}
& \frac{{v({{\textbf{\emph{b}}}^p})}}{{\sum\limits_{j \in supp({{\textbf{\emph{b}}}^p})} {{Q_{p,j}}} }} \le \frac{{v({{\textbf{\emph{b}}}^{p - }})}}{{\sum\limits_{j \in supp({{\textbf{b}}^{p - }})} {{Q_{p,j}}} }} = \frac{{\varphi_p}}{{\rho_p}} \\
\Rightarrow \quad & {p_{p,k}} \le {p_{p,k}}^\prime.
\end{split}
\end{equation}

(iii) $v({{\textbf{\emph{b}}}^p}) = v({\emph{\textbf{b}}^{p - }}) = \varphi_p$
\begin{equation}\label{deduction3}
\begin{split}
& \frac{{v({{\textbf{\emph{b}}}^p})}}{{\sum\limits_{j \in supp({{\textbf{\emph{b}}}^p})} {{Q_{p,j}}} }} < \frac{{v({{\textbf{\emph{b}}}^{p - }})}}{{\sum\limits_{j \in supp({{\textbf{b}}^{p - }})} {{Q_{p,j}}} }} < \frac{{\varphi_p}}{{\rho_p}} \\
\Rightarrow \quad & {p_{p,k}} < {p_{p,k}}^\prime.
\end{split}
\end{equation}

Therefore, we can imply that when user $j$ quits a coalition ${{\textbf{\emph{b}}}^p}$, the utility of other members in this coalition will not be affected, and thus, no users' utility will decrease due to the \emph{transfer operation} of any user $j$. Likewise, the users' utility do not decrease when user $j$ quits a coalition or joins a new coalition, as shown in $\left( \ref{quit_coalition} \right)$ and $\left( \ref{join_coalition} \right)$. Therefore, after iteration ${\ell}$, each user's utility increases or at least remains the same, and the total utility of all the users won't decrease. This also reflects that our proposed algorithm prevents the users from forming an OCS which has previously appeared.

%
%

Given the number of the users and the discrete manner in which the users allocate their resources, the total number of possible OCS is finite. The total utility of all the users has an upper bound due to the limited resources that the users process. Therefore, our proposed algorithm is guaranteed to reach a final OCS within limited iterations. According to \textbf{Lemma 1}, the final OCS resulting from the proposed OCF-algorithm is \emph{T-stable}.
\end{proof}

%

\subsubsection{Complexity}
Note that the total number of iterations cannot be given in the closed form since we do not know for sure at which moment the users form a \emph{T-stable} OCS, which is common in the design of most heuristic algorithms. We will show the distribution of the total number of iterations in Fig.~\ref{iteration_Num_VS_user_Num} and will give corresponding analysis in Section~\textbf{5}.

Note that the computational complexity mainly lies in the number of both the iterations and the attempts of transfer operations in the coalition formation phase. Below we analyze the number of transfer operations in each iteration. We consider the worst case in which each user $j$ joins $\left\lceil {N/2} \right\rceil$ coalitions. In one iteration, each user $j \in \cal{M}$ tries at most $\left\lceil {N/2} \right\rceil \left( {N - \left\lceil {N/2} \right\rceil } \right)$ attempts of \emph{transfer operations}, so the overall number of the attempts in the worst case is $M\left\lceil {N/2} \right\rceil \left( {N - \left\lceil {N/2} \right\rceil } \right)$. In practice, one iteration requires a significantly lower number of attempts, since the number of coalitions that a user joins is usually much smaller than $\left\lceil {N/2} \right\rceil$.

\subsubsection{Signaling Cost}

To describe the signaling cost over the control channels in the proposed OCF-algorithm, we assume that $\eta$ messages are required when a value or the location information of user $j$ is transmitted. We also assume that $\mu$ messages are required when the id of a user or a task is transmitted. In the subcarrier allocation phase, each user $j \in \cal{M}$ needs to report to the platform its id and the location information. Therefore, $MN\eta  + M\mu$ messages are required in the first phase. Here we do not count the messages transmitted between the platform and the BS. In the coalition formation phase, when considering a \emph{transfer operation} ${T_j}({\emph{\textbf{b}}^p},{\emph{\textbf{b}}^q})$, each user $j$ needs to be informed of the members' id and location information\footnote{User $j$ needs to calculate how much payoff it can obtain from coalition $\emph{\textbf{b}}^q$ if it executes \emph{transfer operation} ${T_j}({\emph{\textbf{b}}^p},{\emph{\textbf{b}}^q})$. According to equation $\left( \ref{oneuser_onetask_utility} \right)$ and $\left( \ref{Payoff} \right)$, once user $j$ obtains other members' location information in coalition $\emph{\textbf{b}}^q$, it can know these members' contribution to coalition $\emph{\textbf{b}}^q$, then user $j$'s potential payoff from $\emph{\textbf{b}}^q$ can be obtained.} in coalition ${\emph{\textbf{b}}^q}$, i.e., $\left\{ {{d_{q,k}}\left| {k \in {\cal{M}},g_j^q = 1 \wedge g_k^q = 1} \right.} \right\}$. The platform needs to inform user $j$ of the value of $\rho_i$ and $\varphi_i$. To obtain the upper bound of the signaling cost in one iteration, we consider the worst case in which each user joins $\left\lceil {N/2} \right\rceil$ coalitions in one iteration. In this case, the signaling cost of all the users is $M\left\lceil {N/2} \right\rceil \left( {N - \left\lceil {N/2} \right\rceil } \right)\left[ {\left( {M + 1} \right)\eta  + \left( {M - 1} \right)\mu } \right]$ messages. Besides, after the users decide which coalitions to participate in, they need to report to the platform the id of their chosen coalitions, and the platform offers corresponding rewards to them, which requires at most $2MN\mu$ messages.

Note that the signaling cost of the proposed OCF-algorithm increases quickly with both the number of coalitions and the size of each coalition. Nevertheless, the size of coalitions is constrained by the limited subcarrier resources that each user has access to, and the number of coalitions equals to the number of tasks. Therefore, the signaling cost of the proposed OCF-algorithm can be restricted to a tolerable level.

\section{Simulation Results}%

To evaluate the performance of our proposed algorithms, we use a simulation setup as follows. We consider a 10km$ \times 10$km square area where the tasks and the users are randomly located. The BS is located at the centre of this square area. The number of the subcarriers is 60. We set the average transmission power of a smartphone $P_u = 23$dBm, noise variance of the transmission channel ${\sigma ^2} = -90$dBm, the bandwith of each subchannel $B = 15$kHz, scale factor $\beta  = 7, \gamma = 0.2$ and the exponential factor $\lambda  = 0.8$. To depict the difference of the tasks, we assume that the upper bound of any task $i$'s sensing performance $\varphi_i$ is uniformly distributed within a range of $90 \sim 150$. Similarly, the threshold of a user's effective contribution to task $i$, $\rho_i$, is uniformly distributed within a range of $35 \sim 60$. The scaling factor ${a_i}$ is uniformly distributed within a range of $3 \sim 7$. The data transmission rate required by task $i$, ${r_i}$, is uniformly distributed within a range of $6 \sim 12$Mb/s, and the radius of task $i$'s AoI, ${{d_{i,0}}}$ is uniformly distributed within a range of $0.6 \sim 2.5$km. All the curves are generated based on averaging over 1000 instances of the algorithms.

\begin{figure}[!t]
\centering
\includegraphics[width=5in]{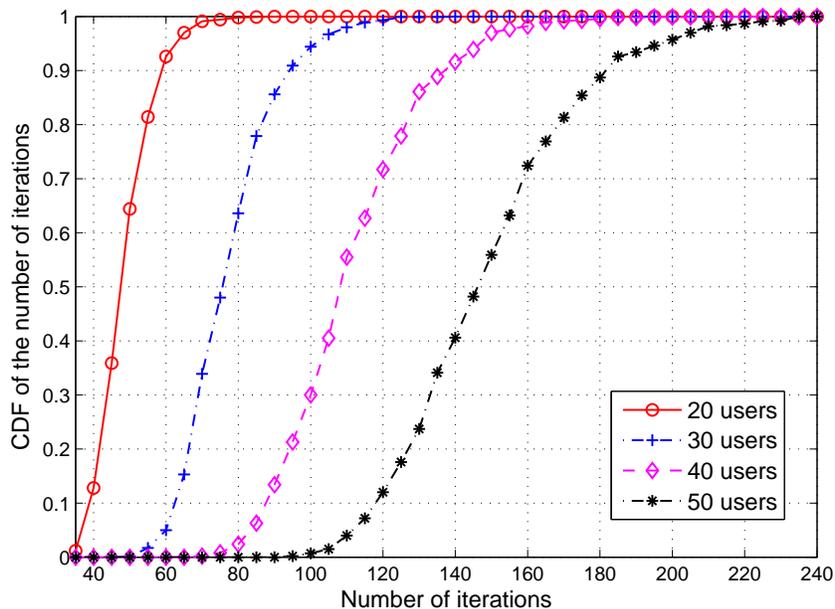}
\caption{CDF of the total number of iterations in the OCF-algorithm versus the number of iterations with the number of the tasks $N = 20$.} \label{iteration_Num_VS_user_Num}
\end{figure}

Denote the random variable $\tilde Y$ as the total number of iterations required for the OCF-algorithm to converge. Fig.~\ref{iteration_Num_VS_user_Num} shows the cumulative distribution function (c.d.f.) of $\tilde Y$, $\Pr \left( {\tilde Y \le \tilde y} \right)$, versus $\tilde y$ for different number of users, with the number of the tasks as 30. We observe that the speed of convergence becomes faster as the number of the users decreases. Fig.~\ref{iteration_Num_VS_user_Num} further reflects that the computational complexity is rather low in our proposed algorithm. For example, when the number of the users is 20, we observe that on average a maximum of only 80 iterations are needed for the OCF-algorithm convergence.

\begin{table}
\centering
\caption{Number of iterations in the OCF-algorithm and the non-cooperative algorithm} \label{complexity}
\begin{tabular}{|m{33mm}|m{15mm}|m{20mm}|m{20mm}|m{18mm}|}
\hline Number of the users (20 tasks)& 20 & 30 & 40 & 50\\
\hline Number of iterations using OCF-algorithm & 80 & 120 & 170 & 240\\
\hline Number of iterations using Non-cooperative algorithm & over $10^6$ & over $2.5 \times {10^6}$ & over $4 \times {10^6}$ & over $5 \times {10^6}$\\
\hline
\end{tabular}
\end{table}

\begin{figure}
\centering
\subfigure[Non-cooperative approach]{
\label{alphatwo:a} 
\includegraphics[width=3.29in]{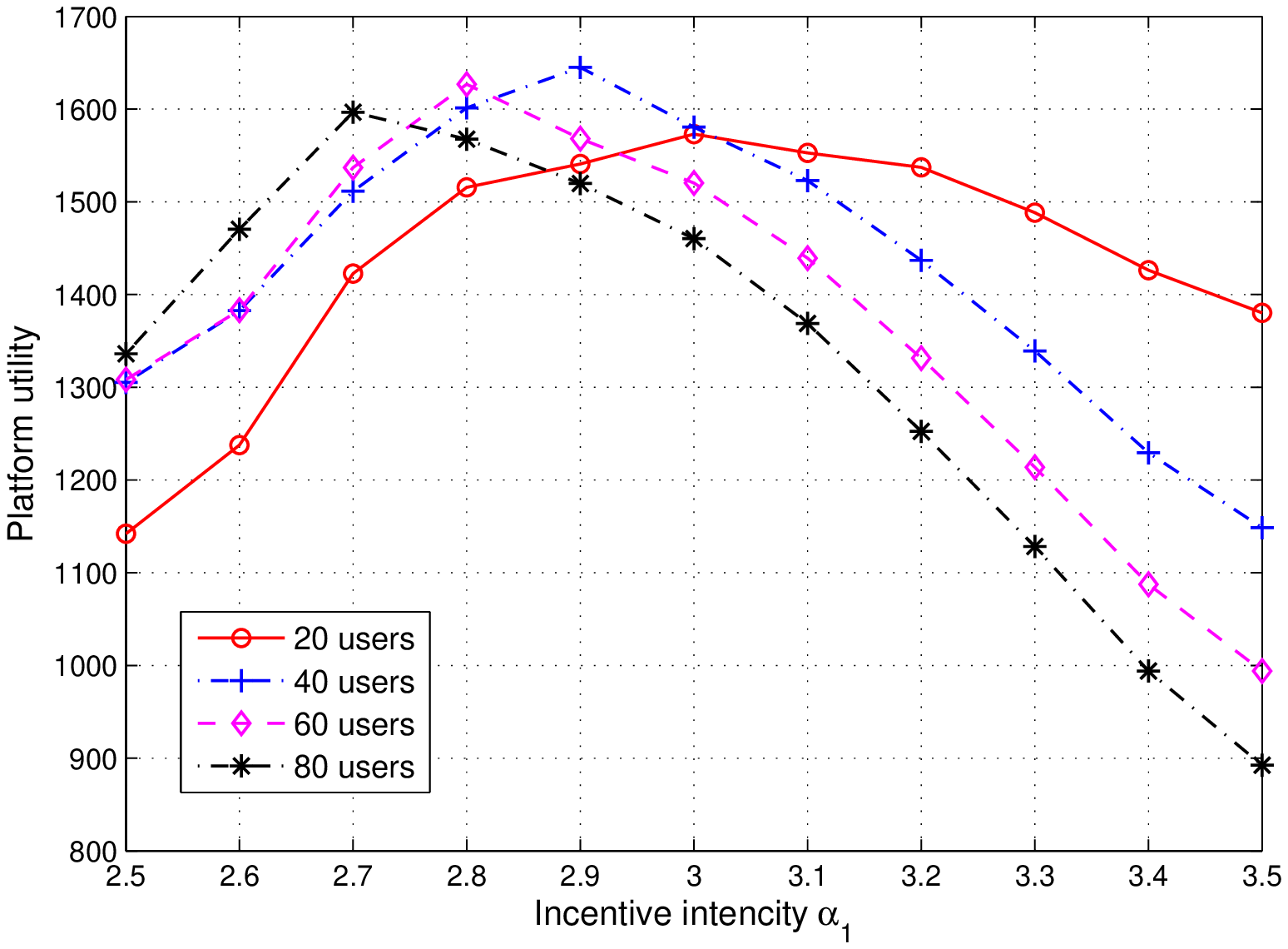}}
\hspace{-0.3in}
\subfigure[Cooperative approach]{
\label{alphatwo:b} 
\includegraphics[width=3.29in]{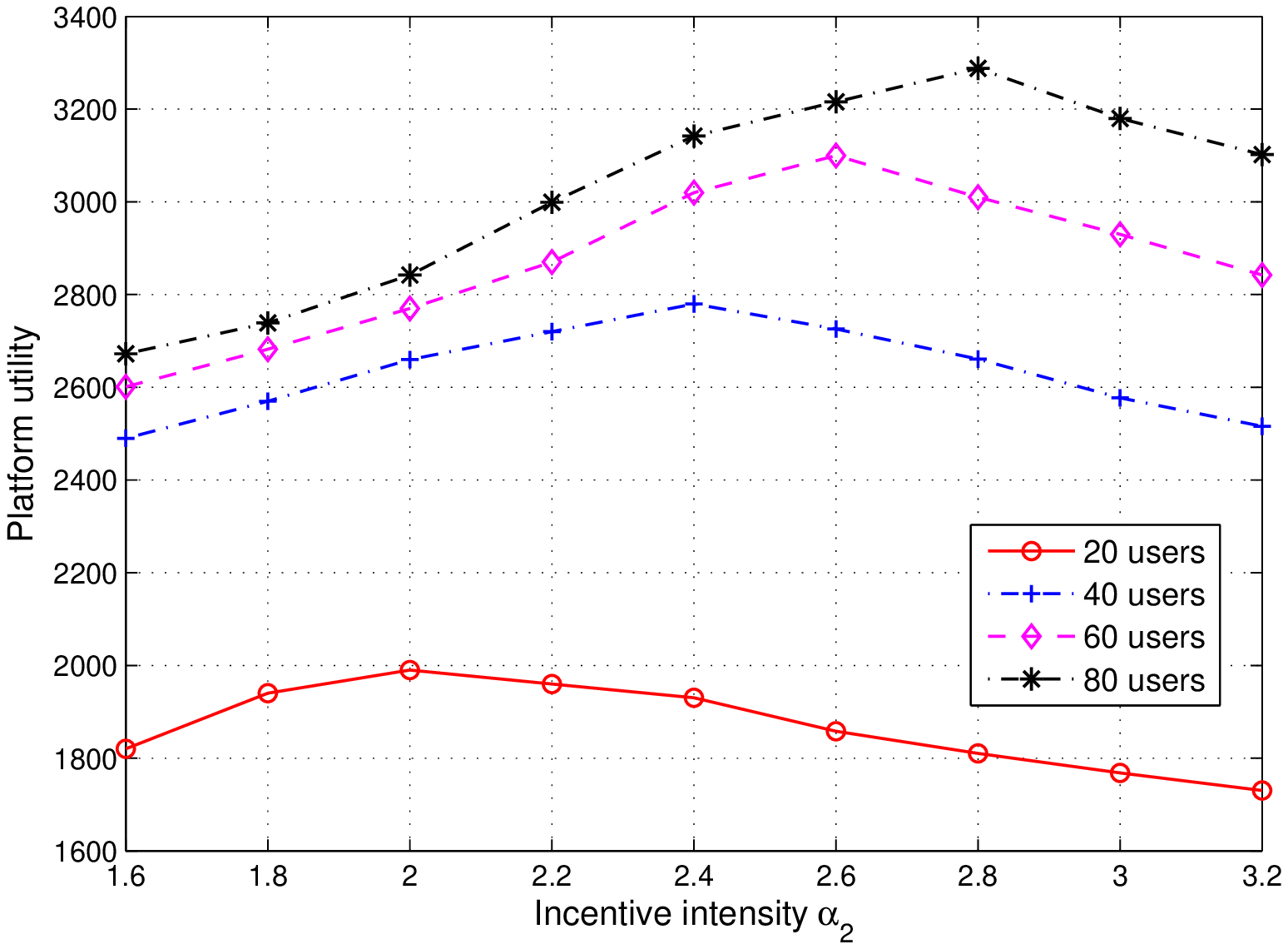}}
\caption{The platform utility effected by the value of the incentive intensity ${\alpha_1}$ in the non-cooperative approach, and by the value of the incentive intensity $\alpha_2$ in the cooperative approach, with the number of the tasks $N = 30$ } \label{alphatwo} 
\end{figure}
\begin{figure}[!t]
\centering
\includegraphics[width=5in]{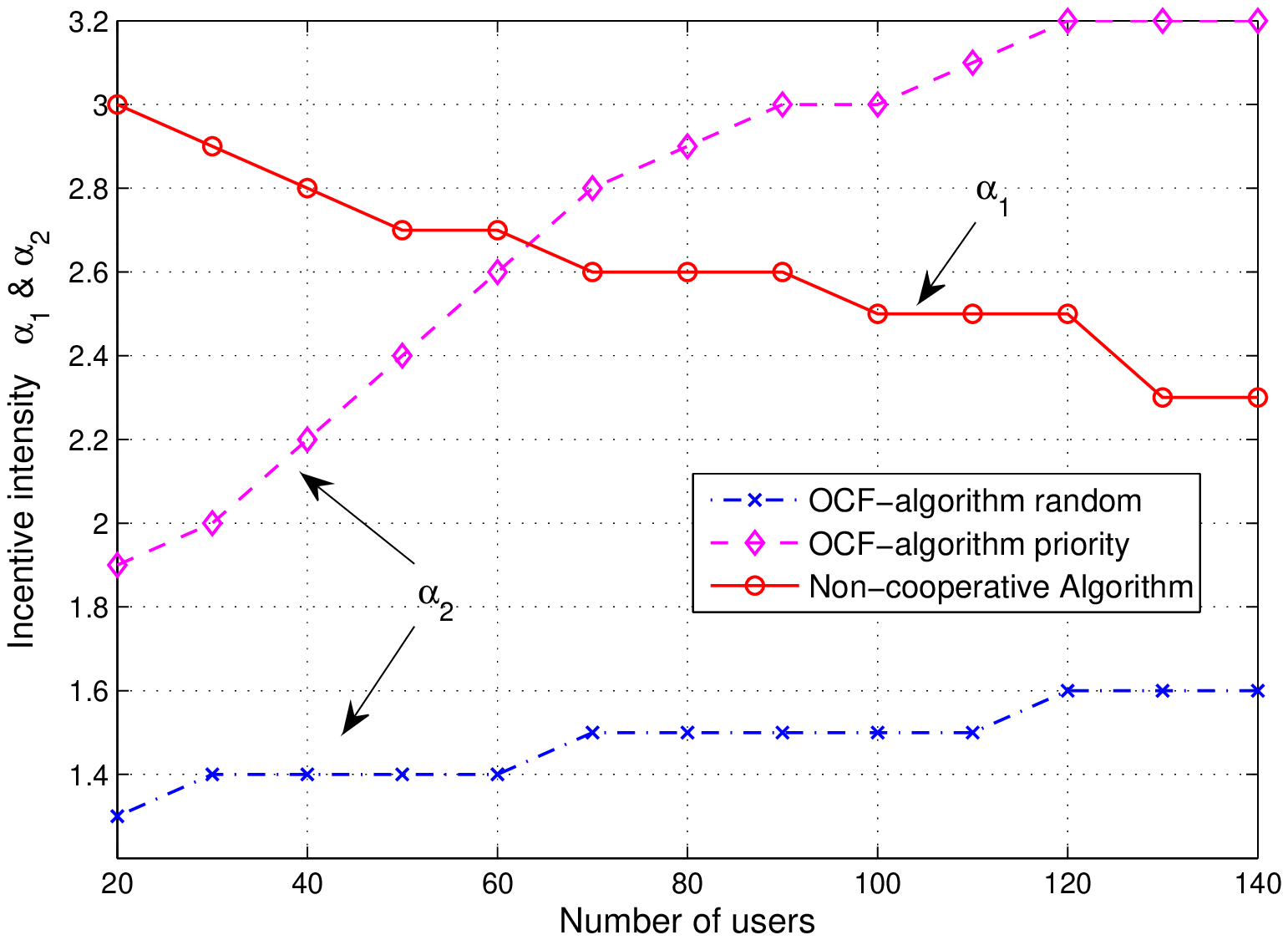}
\caption{The best values of the incentive intensity $\alpha_1$ and $\alpha_2$ effected by the number of users in the non-cooperative approach and the cooperative approach with the number of the tasks $N = 30$.} \label{user_num_four_alpha}
\end{figure}

\begin{figure}[!t]
\centering
\includegraphics[width=5in]{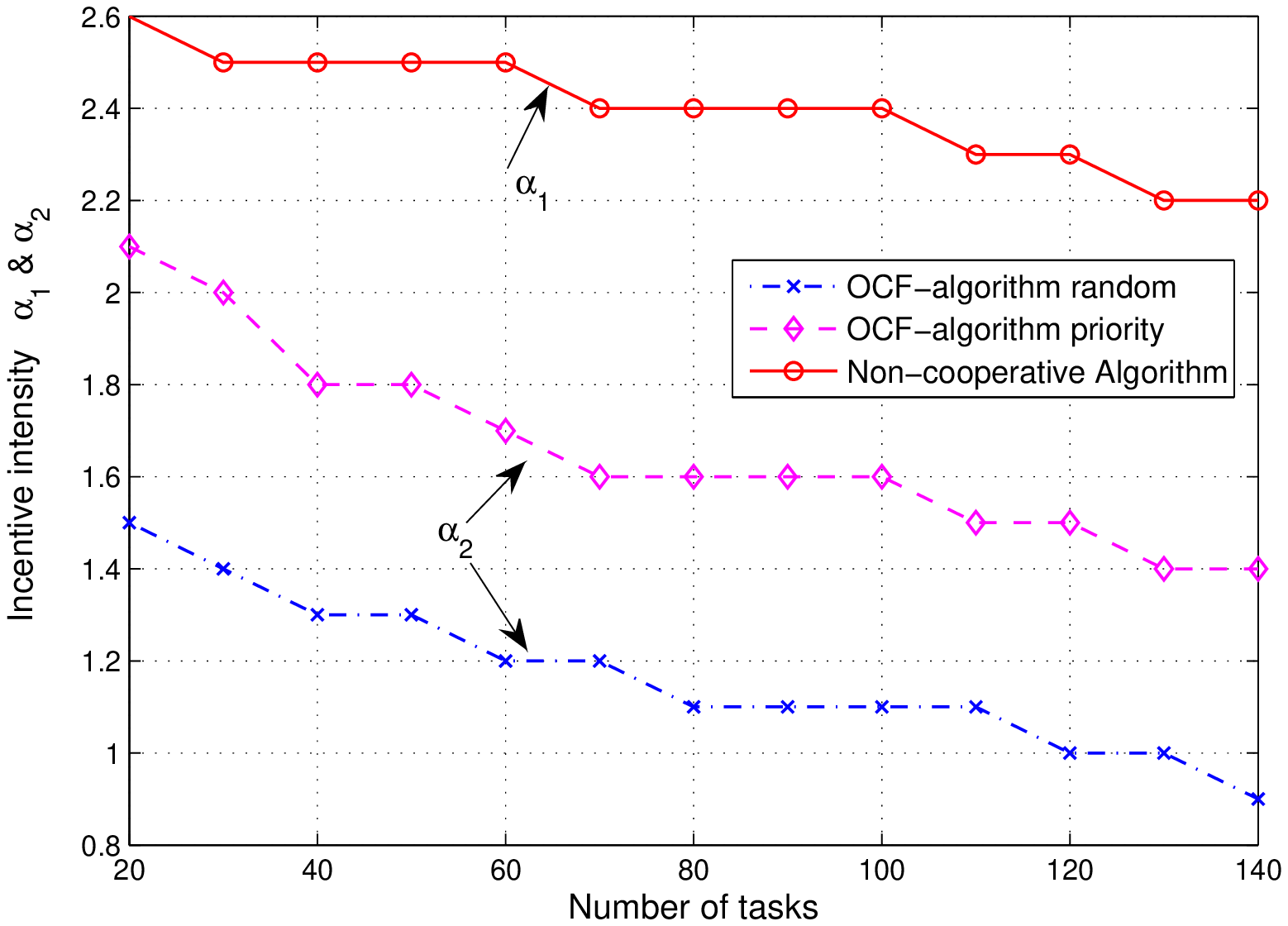}
\caption{The best values of the incentive intensity ${\alpha_1}$ and ${\alpha_2}$ effected by the number of tasks in the non-cooperative approach and the cooperative approach with the number of the users $M = 80$.} \label{task_num_four_alpha}
\end{figure}

\begin{figure}[!t]
\centering
\includegraphics[width=5in]{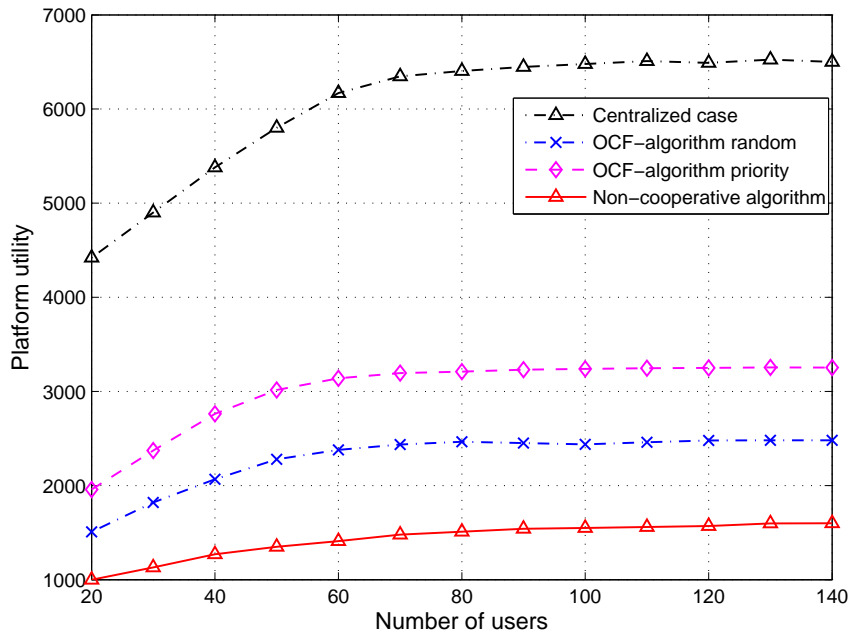}
\caption{Impact of the number of users to platform utility with the number of the tasks $N = 25$.} \label{platform_utility_user_num}
\end{figure}

\begin{figure}[!t]
\centering
\includegraphics[width=5in]{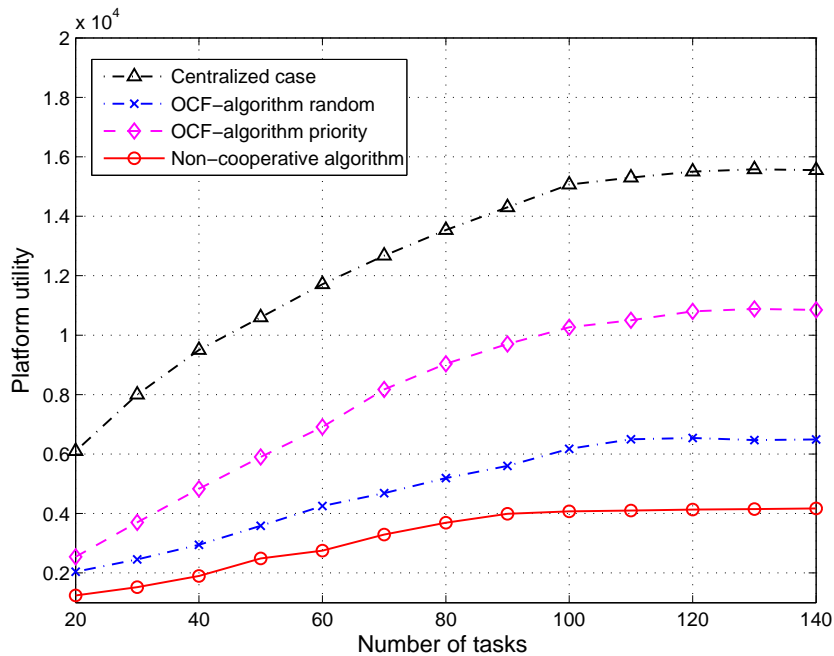}
\caption{Platform utility effected by the number of tasks with the number of the users $M = 80$.} \label{task_num_platform_utility}
\end{figure}

\begin{figure}[!t]
\centering
\includegraphics[width=5in]{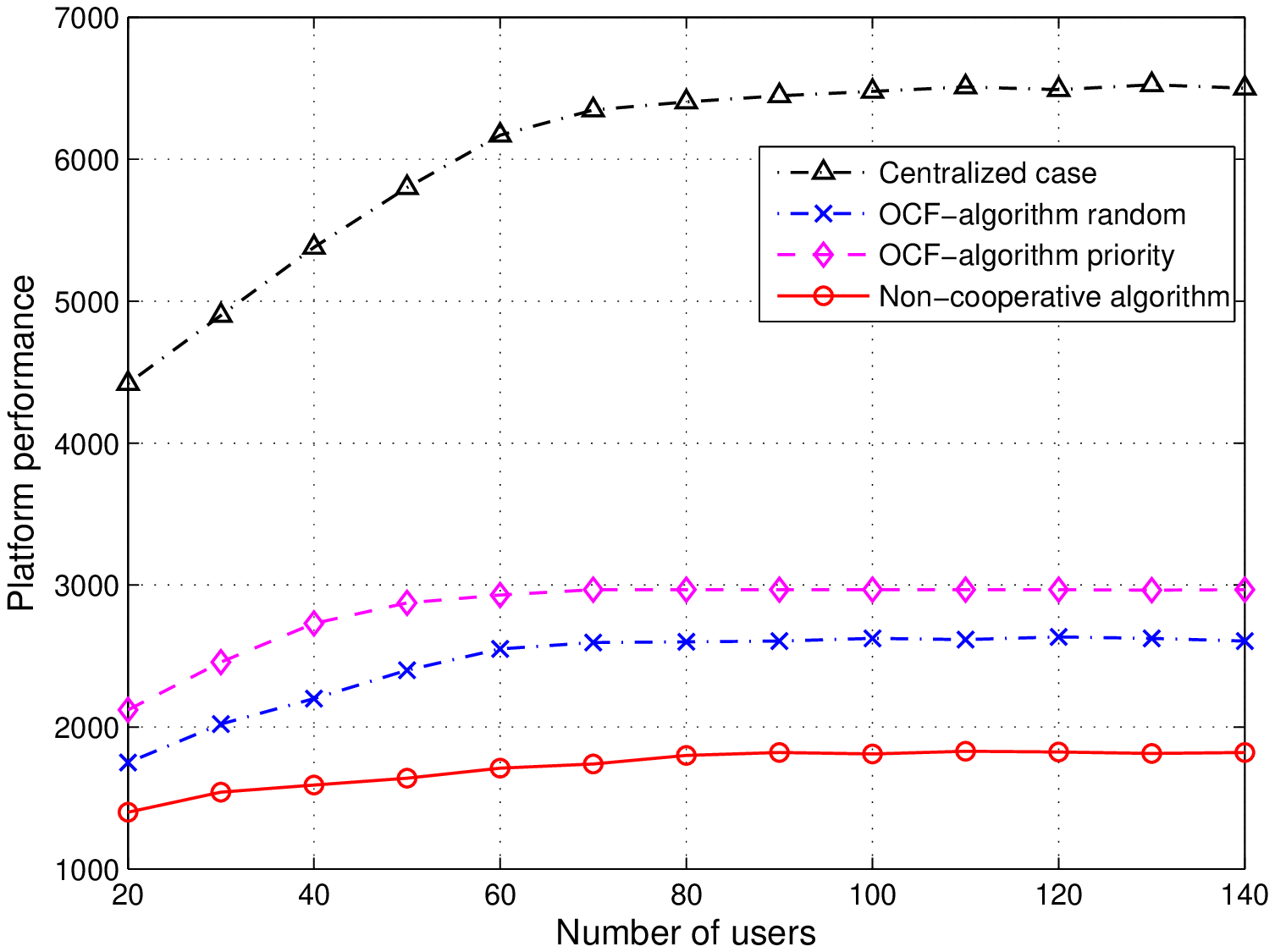}
\caption{The platform performance effected by the number of users with the number of the tasks $N = 25$.} \label{user_num_platform_performance}
\end{figure}

In Table~\ref{complexity}, we show the number of iterations in both the OCF-algorithm and the non-cooperative algorithm with the number of tasks set as 20. For the non-cooperative algorithm, we give the number of iterations when solving the 0-1 INLP problem in $\left( \ref{noncooperative_utility_maximazition} \right)$ with optimization tolerance set to 0.01. For the OCF-algorithm, we set the number of iterations as the worst-case value. From this table, we can see that the computational complexity of the non-cooperative approach is extremely high compared to that of the OCF-algorithm.

Fig.~3 shows the platform utility in the non-cooperative approach as a function of the value of ${\alpha_1}$, and the platform utility in the cooperative approach as a function of the value of ${\alpha_2}$. From the curves in Fig.~3a, we see that there exists a best value of ${\alpha_1}$ which makes the platform utility reach the peak value. As mentioned in Section \textbf{2.2}, the value of ${\alpha_1}$ reflects the degree how the platform motivates the users. When ${\alpha_1}$ is too small, the incentive mechanism does not work effectively since the users are not willing to participate in the tasks with such a low payoff. When ${\alpha_1}$ is too large, the incentive cost is so much for the platform that it affects the platform utility though the users are well motivated. Therefore, there must exist a value of ${\alpha_1}$ that achieves the highest platform utility, i.e., the trade-off between the incentive cost and the incentive effect can be achieved. Similar analysis stands for the curves of the cooperative approach in Fig.~3b. In both subgraphs, the best incentive intensity changes as the number of users changes.

Fig.~\ref{user_num_four_alpha} shows the best values of $\alpha_1$ and $\alpha_2$ as a function of the number of users in both the non-cooperative approach and the cooperative approach. From Fig.~\ref{user_num_four_alpha}, the best value of ${\alpha_1}$ decreases as the number of users grows in the non-cooperative approach, while the best value of $\alpha_2$ increases as the number of users grows in the cooperative approach. This is because in the non-cooperative approach, as the users are becoming more, the competition between the users participating in the tasks is more intense. However, there is no improvement in their utilization of the subcarrier resources, and thus, the platform needs to cut its incentive cost so to maintain as much platform utility as possible. In the cooperative approach, as the number of the users increases, the users can utilize the subcarrier resources more efficiently by cooperating with each other, thus there is still a rising space for the platform utility if the incentive intensity is larger.

Fig.~\ref{task_num_four_alpha} shows the best values of $\alpha_1$ and $\alpha_2$ as a function of the number of tasks in both the non-cooperative approach and the cooperative approach with the number of the users set as 80. The best values of ${\alpha_1}$ and ${\alpha_2}$ decease as the number of tasks increases, which means that the best incentive intensity responses the same to the change of the number of the tasks in both approaches. This is because when there are more tasks in the network, the users are more selective about the tasks to participate in, so both the competition in the non-cooperative approach and the cooperation in the cooperative approach are weakened. Therefore, the users themselves are more inclined to participate in the sensing tasks since there is more opportunity for them now to gain high profits from those newly publicized tasks, which leads to the decrease of the platform's incentive cost.

Fig.~\ref{platform_utility_user_num} shows the platform utility as a function of the number of users with the number of the tasks set as 25. The optimal values of $\alpha_1$ and $\alpha_2$ are adopted in the algorithms separately. The platform utility increases with the number of users, and turns out to be diminishing returns, which can be explained as below. As the number of users increases, the platform can recruit more users, thereby improving the social welfare. However, the number of the subcarriers is limited to 60 and one subcarrier can only be assigned to a single user, so at most 60 users have access to the tasks. Besides, as the number of users grows, the contribution they make to the platform become saturated, and thus, the platform utility increases to reach a stable value, which leads to the diminishing returns of the curves. In addition, Fig.~\ref{platform_utility_user_num} shows the upper bound of the platform utility, which is represented by the result of the centralized case, where the users get no payoff from the platform. The OCF-algorithm performs better than the non-cooperative approach since the waste of resources is avoided in the OCF-algorithm, which helps improve the sensing performance.

Fig.~\ref{task_num_platform_utility} shows the platform utility as a function of the number of tasks with the number of the users set as 80. We observe that as the platform publicizes more tasks, the platform utility increases and converges to a maximum value. When there are too many tasks in the system, the users are not able to participate in every task due to their limited subcarrier resources, and thus, there is always a point at which the platform utility reaches a maximum value. Besides, we can see that the cooperative approach performs better than the non-cooperative one.

Fig.~\ref{user_num_platform_performance} shows the platform sensing performance as a function of the number of users with the number of tasks set as 25. Note that the platform's sensing performance directly reflects the effects of different algorithms. The trends of all the methods are similar with those of the platform utility, but the gaps between various methods are narrowed, because the differences of incentive cost are not considered here.

\section{Conclusions}%

In this paper, we have proposed two incentive mechanisms applying the non-cooperative approach and the cooperative approach, respectively, for the smartphone sensing system. There are four factors that influence the platform utility, which are the numbers of users, tasks and subcarriers, and the incentive intensity of the mechanism. The numbers of users, tasks and subcarriers are three factors restricted with each other, i.e., with two of these factors fixed, the platform utility increases and flattens out as the third one grows. The incentive intensity of the mechanism in the non-cooperative approach and the cooperative approach,  i.e., $\alpha_1$ and $\alpha_2$, respectively, greatly influences the platform utility by affecting the users' behaviors. A trade-off between the incentive cost and the incentive effect of the platform can be achieved by trying different values of the incentive intensity. There exists the best value of the incentive intensity in both approaches. From the simulation results, when the best values of $\alpha_1$ and $\alpha_2$ are adopted in the non-cooperative approach and the cooperative approach, respectively, the latter one performs better than the former one in terms of both the platform utility and the sensing performance.



\end{document}